\documentclass[preprint,11pt]{elsarticle}

\usepackage{graphicx}
\usepackage{tabularx}
\usepackage{amssymb}
\usepackage{amsmath}
\DeclareMathOperator*{\argmax}{argmax} 
\usepackage{bm}
\usepackage{commath}
\usepackage{xcolor}

\usepackage{epstopdf}
\AppendGraphicsExtensions{.tif}

\usepackage{caption,booktabs}
\usepackage{array}
\usepackage{multirow}
\newcolumntype{P}[1]{>{\centering\arraybackslash}p{#1}}
\usepackage{enumitem}

\journal{}
\date{}

\begin{document}
	\begin{frontmatter}
		
		
		\title{Data-Driven Modeling of Coarse Mesh Turbulence for Reactor Transient Analysis Using Convolutional Recurrent Neural Networks}		

		
		\author[1]{Yang Liu}
		\author[1]{Rui Hu}
		\author[1,2]{Adam Kraus}
		\author[3]{Prasanna Balaprakash}
		\author[4]{and Aleksandr Obabko}
		\address[1]{Nuclear Science and Engineering Division, Argonne National Laboratory}
		\address[2]{Ken and Mary Alice Lindquist Department of Nuclear Engineering, Penn State University}
		\address[3]{Mathematics and Computer Science Division, Argonne National Laboratory}
		\address[4]{Computational Science Division, Argonne National Laboratory}
		
		\begin{abstract}
			Advanced nuclear reactors often exhibit complex thermal-fluid phenomena during transients. To accurately capture such phenomena, a coarse-mesh three-dimensional (3-D) modeling capability is desired for modern nuclear-system code. In the coarse-mesh 3-D modeling of advanced-reactor transients that involve flow and heat transfer, accurately predicting the turbulent viscosity is a challenging task that requires an accurate and computationally efficient model to capture the unresolved fine-scale turbulence.\\
			In this paper, we propose a data-driven coarse-mesh turbulence model based on local flow features for the transient analysis of thermal mixing and stratification in a sodium-cooled fast reactor. The model has a coarse-mesh setup to ensure computational efficiency, while it is trained by fine-mesh computational fluid dynamics (CFD) data to ensure accuracy. A novel neural network architecture, combining a densely connected convolutional network and a long-short-term-memory network, is developed that can efficiently learn from the spatial-temporal CFD transient simulation results. The neural network model was trained and optimized on a loss-of-flow transient and demonstrated high accuracy in predicting the turbulent viscosity field during the whole transient. The trained model’s generalization capability was also investigated on two other transients with different inlet conditions. The study demonstrates the potential of applying the proposed data-driven approach to support the coarse-mesh multi-dimensional modeling of advanced reactors.  
		\end{abstract}
	

	\end{frontmatter}
	
	
	\section{Introduction}
	\label{S:1}
    Advanced reactors are expected to fulfill a key role in next-generation nuclear power plants because of their increased safety and economic performance. Leveraging the full potential of these new reactor technologies requires modern analysis tools to ensure the safety of the new designs. As an example, the System Analysis Module (SAM), a modern system-level code, is currently under active development at Argonne National Laboratory for advanced reactor design and safety analysis \cite{hu2021sam}.
    
    Although the main objective of a system-level code is to conduct whole-plant transient analysis, three-dimensional (3-D) simulation capability is still desirable to tackle complex thermal-fluid (T-F) phenomena in advanced reactors. A typical example is the mixing and thermal stratification phenomena in large pools or enclosures. Such phenomena can be observed in various reactor systems, including the cold- and hot-pool mixing in pool-type sodium-cooled fast reactors \cite{lu2020sensitivity}, reactor cavity cooling system behavior in high-temperature gas-cooled reactors \cite{gou2018preliminary}, passive containment cooling in advanced light-water reactors \cite{tang2021numerical}, and thermal stratification in boiling-water reactor suppression pools \cite{bao2018safe}.
    
    It is very important to accurately predict pool temperature and density distributions for both design optimizations and safety analyses of these reactor systems. However, the individual transport mechanisms governing mixing are characterized by time and length scales that can differ by orders of magnitude. Large volumes and complex interactions of different flow and thermal structures make the analysis of mixing in a large enclosure a very challenging task. Current major reactor system analysis codes have either no models or only coarse one-dimensional (1-D) models for thermal mixing and stratification in large enclosures. The lack of general thermal mixing and stratification models in those codes severely limits their application and accuracy for safety analysis, especially for reactors relying on natural circulation for long-term cooling.
    
    For this reason, a 3-D module is currently being developed in SAM to accurately model complex T-F phenomena while maintaining the computational efficiency of system code \cite{hu2019three}. This module adopts a coarse-mesh setup to be consistent with the one-dimensional system modeling framework, which ensures computational efficiency. A major challenge for the coarse-mesh 3-D module in SAM is to accurately capture turbulence \cite{zou2020development} in reactor transients. In computational fluid dynamics (CFD), the high-fidelity simulations, such as direct numerical simulation (DNS) or large eddy simulation (LES), are computationally expensive and are not widely used for modeling long reactor transients. A widely used approach for turbulence prediction in such engineering applications is the two-equation turbulence model based on the Reynolds-averaged Navier-Stokes (RANS) equations. However, this approach still requires a fine-mesh setup, so it is incompatible with the coarse-mesh 3-D module in system code such as SAM. 
    
    Data-driven approaches enabled by machine learning (ML) have led to significant progress in many nuclear engineering applications, including reactor performance optimization \cite{radaideh2021physics, wang2021moisture, sun2020optimizing}, transient-state prediction \cite{guillen2020relap5, ahn2020deep}, anomaly detection \cite{kim2020rnn}, data assimilation \cite{gong2020inverse, gong2021optimal}, model validation and uncertainty quantification \cite{wu2018inverse_1, wu2018inverse_2, liu2019uncertainty, radaideh2019integrated, liu2019validation}, and digital twin \cite{lin2021dt}. Among various ML methods, deep neural networks (DNNs) provide us with an especially promising technical approach to developing a data-driven coarse-mesh turbulence model. DNNs have demonstrated promising results \cite{iskhakov2021integration, raissi2019physics, long2019pde} in achieving approximate solutions of partial differential equations (PDEs), with two advantages: First, DNNs are capable of modeling nonlinear relationships between inputs and outputs, as proven by the universal approximation theorems \cite{hornik1991approximation}. Second, once the DNN is trained, its forward evaluation is very fast, making it computationally preferable to introducing additional PDEs to the system code for turbulence prediction. These advantages make the DNN-based data-driven turbulence model possible for nuclear-system code. Several recent applications to T-F modeling using DNNs in the nuclear engineering community further demonstrate the applicability of DNNs \cite{liu2018data, bao2019data, liu2020coarse, liu2021uncertainty,liu2021dnn_uq}. Leveraging DNNs, we can learn from the high-resolution CFD data to develop coarse mesh model that is able to retain the accuracy of the CFD results. Such an approach is backed by the progresses in high-fidelity CFD simulations on complex nuclear reactor components, from fuel assembly to reactor pool \cite{kraus2021large, buzzi2020analysis, colombo2021study, huning2021review}.
    
    In this paper, we propose to leverage DNN to develop a data-driven coarse-mesh turbulence model for modeling reactor transients that involve thermal mixing/stratification in large open volumes. The DNN model is trained with fine-mesh CFD results to ensure accuracy, while it has a coarse-mesh setup so its computational efficiency is similar to that of system code. We adopted the convolutional recurrent neural network architecture to capture the spatial-temporal information from CFD simulation results during reactor transients. We performed a case study on a simplified Sodium Fast Reactor (SFR) hot pool with several transient simulations to demonstrate the applicability of the developed DNN model. The developed model can be used for modern system code with multi-dimensional modeling capability, such as SAM.
    
    The rest of this paper is organized as follows: Section \ref{S:2} introduces the proposed neural network; Section \ref{S:3} discusses the reactor transients we investigated in this paper; Section \ref{S:4} provides details on the case study; Section \ref{S:5} evaluates the performance of the trained neural network model; and Section \ref{S:6} provides summary remarks. 
    
   \section{Deep neural network for coarse-mesh turbulence prediction}
   \label{S:2}
   
   \subsection{Progress on data-driven turbulence modeling}
   As an active research area, data-driven turbulence modeling based on ML methods has achieved significant progresses in the past few years \cite{duraisamy2019turbulence, xiao2019quantification, duraisamy2021perspectives}. Generally speaking, these works can be characterized into two categories: a). ML models serve as the error correction term that will be added to a certain term in the turbulence model, such as the Reynolds stress term in RANS, or the subgrid-scale term in LES; b) ML models serve as full data-driven closures to replace the corresponding terms in the turbulence model.
   
   In the error correction approach, Duraisamy et al. \cite{parish2016paradigm, singh2016using} used field inversion to infer multiplier terms in the RANS turbulence closures. Xiao et al. \cite{xiao2016quantifying} used Kalman filtering method to inversely train Gaussian processes to represent the model discrepancy of Reynolds stress for RANS turbulence model. Similarly, Wang et al. \cite{wang2017physics} used random forest algorithm to reconstruct Reynolds stress discrepancies for RANS turbulence model. Beck et al. \cite{beck2019deep} used residual neural network to train spatial terms for LES equations.
   
   In the full data-driven closure approach, a pioneering work was done by Ling et al. \cite{ling2016reynolds}. In this work, a tensor basis neural network that embed Galilean invariance is developed to predict the anisotropic part of the Reynolds stress tensor. A later work done by Milani et al. \cite{milani2021generality} also showed the proposed tensor basis network model have reasonable generality on turbulent scalar flux modeling. Wang et al. \cite{wang2018investigations} used multi-layer perceptron (MLP) to develop closure model of subgrid-scale stress for LES. In the work, influence of input features on the ML model was investigated, and they chose the first- and second-order partial derivatives of velocity components as inputs for their ML model. Maulik et al. \cite{maulik2018data} used neural networks for the convolution and deconvolution of coarse-mesh fields to account for subgrid-scale turbulence effects. They showed that turbulent-viscosity can be accurately predicted in 2-dimensional Kraichnan turbulence with the pure data-driven approach. On the other hand, Wu et al. \cite{wu2019reynolds} revealed that RANS equations with an explicit data-driven Reynolds stress closure can be ill-conditioned.
   
   It should be noted that for both above-mentioned approaches, the ML models were trained with high-fidelity DNS/LES data on statistically stationary simulations and the goal is to improve RANS or LES turbulence model. The focus is on addressing some fine-scale turbulence effects such as the anisotropic part of the Reynolds stress in RANS or the subgrid-scale closure in LES.
   
   \subsection{Design a data-driven turbulence model for reactor transient analysis}
   The goal of this work is to develop a coarse-mesh closure for the isotropic turbulent viscosity to support the safety modeling of advanced nuclear reactors, which often involve reactor transients with minutes or hours long with dynamic changes in inlet and boundary conditions. In this sense, the problem is spatial-temporal dynamic in nature: the computation domain is multi-dimensional, and the reactor transients involve dynamic changes in inlet and boundary conditions over time. We consider the previous works for full data-driven closures discussed in Section 2.1 not directly applicable for this goal for two reasons. Firstly, it is impractical to obtain high-fidelity DNS/LES results for such long transients, and RANS is the only available approach to provide training data to train the data-driven turbulence closure. Secondly, in nuclear system code, we do not aim to resolve fine-scale turbulence effect such as the anisotropic part of Reynolds stress. Instead, the desired coarse-mesh turbulence model is the isotropic turbulence in a much coarser scale so that the corresponding multi-dimensional module can be consistent with the one-dimensional components of nuclear system code.
    
   The desired DNN model for this work should be a field-to-field mapping that can be efficiently implemented into nuclear system code. In this work, we chose be a field-to-field mapping that takes the local flow feature fields
   $\bm{Q}_t=[\bm{q}_t^\mathit{1},\bm{q}_t^\mathit{2},\cdots,\bm{q}_t^\mathit{n}]$ as inputs and is informed by previous time-step information $\mathcal{L}_{t-1}$ to predict the turbulent viscosity field $ \bm{\mu}_t^t $ at time step $ \mathit{t} $:
   
   \begin{equation}
   \label{eq:1}
   f\left(\bm{Q}_t,{\ \mathcal{L}}_{t-1}\right)=\ \bm{\mu}_t^t,
   \end{equation}
   
   \noindent
    where $ \bm{q}\in\mathbb{R}^{X\times Y\times Z} $ is the flow-feature field of the computation domain with the dimensionality of $ X\times Y\times Z $(depending on the mesh size),  $ \bm{\mu}_t^t $ is the turbulent viscosity field with the same dimensionality, and $\mathcal{L}_{t-1}$ is the latent space information of previous time steps with a compressed dimensionality.

   Convolutional neural networks (CNNs) \cite{lecun1995convolutional} have demonstrated superior performance in the data-driven modeling of spatial-temporal dynamic systems \cite{xingjian2015convolutional, xu2020multi, maulik2021reduced}. For this reason, the backbone of the field-to-field mapping is chosen to be CNN, which is also informed by the recurrent neural network (RNN) to learn the temporal information.

   There are additional advantages for developing the mapping based on CNN. First, convolution can serve as a numerical approximator of differential operators, as Dong et al. \cite{dong2017image, long2019pde} proved that convolution operation has a profound mathematical relationship with differentiation operators. In this sense, CNN can be used to implicitly learn derivatives of an input physical quantity, so these derivative terms are not needed to be explicitly included as inputs. Such an adoption significantly simplifies the problem.

   Furthermore, the RANS equations that describe the thermal-fluid system obey Galilean invariance, which includes rotation invariance, translation invariance, and uniform motion invariance. In this work, the goal is to model the isotropic turbulent viscosity, so the rotational invariance is insignificant under the isotropic assumption. The derivative terms of physical quantities satisfy the invariance of translation and uniform motion \cite{wang2018investigations}. So in the context of this work, the CNN architecture is consistent with the Galilean invariance.
   
   \subsection{DCNN-LSTM based on convolutional recurrent neural network}
   CNN is based on multiple convolutional kernels that slide along the input fields to produce lower-dimensional feature maps. The convolution operation on fields $\bm{Q}$ with kernel $\bm{K}$ can be expressed as follows:
     
   \begin{equation}
   \label{eq:2}
   (\bm{Q}\ast\bm{K})\left(x,y,z\right)=\sum_{m}\sum_{n}\sum_{l}{\bm{Q}\left(x-m,\ y-n,\ z-l\right)\bm{K}(m,n,l)}.
   \end{equation}
   
   The convolution product depends on the size (denoted by $\mathit{k}$) and the weights of the kernel, as well as the convolution operation’s stride $\mathit{s}$ and padding $\mathit{p}$. 
   
   Compared to feedforward neural networks, the weight-sharing and sparse connection features of CNN are advantageous for learning from 3-D data. Furthermore, previous studies reveal that the convolution operation has a profound mathematical relationship with differentiation operators \cite{dong2017image}. Supported by this relationship, we do not need to extract gradients of flow fields as inputs when training the neural network; this feature significantly simplifies the problem compared to the classical feedforward neural-network approach.
   
\begin{figure}[h]
   	\centering\includegraphics[width=0.7\linewidth]{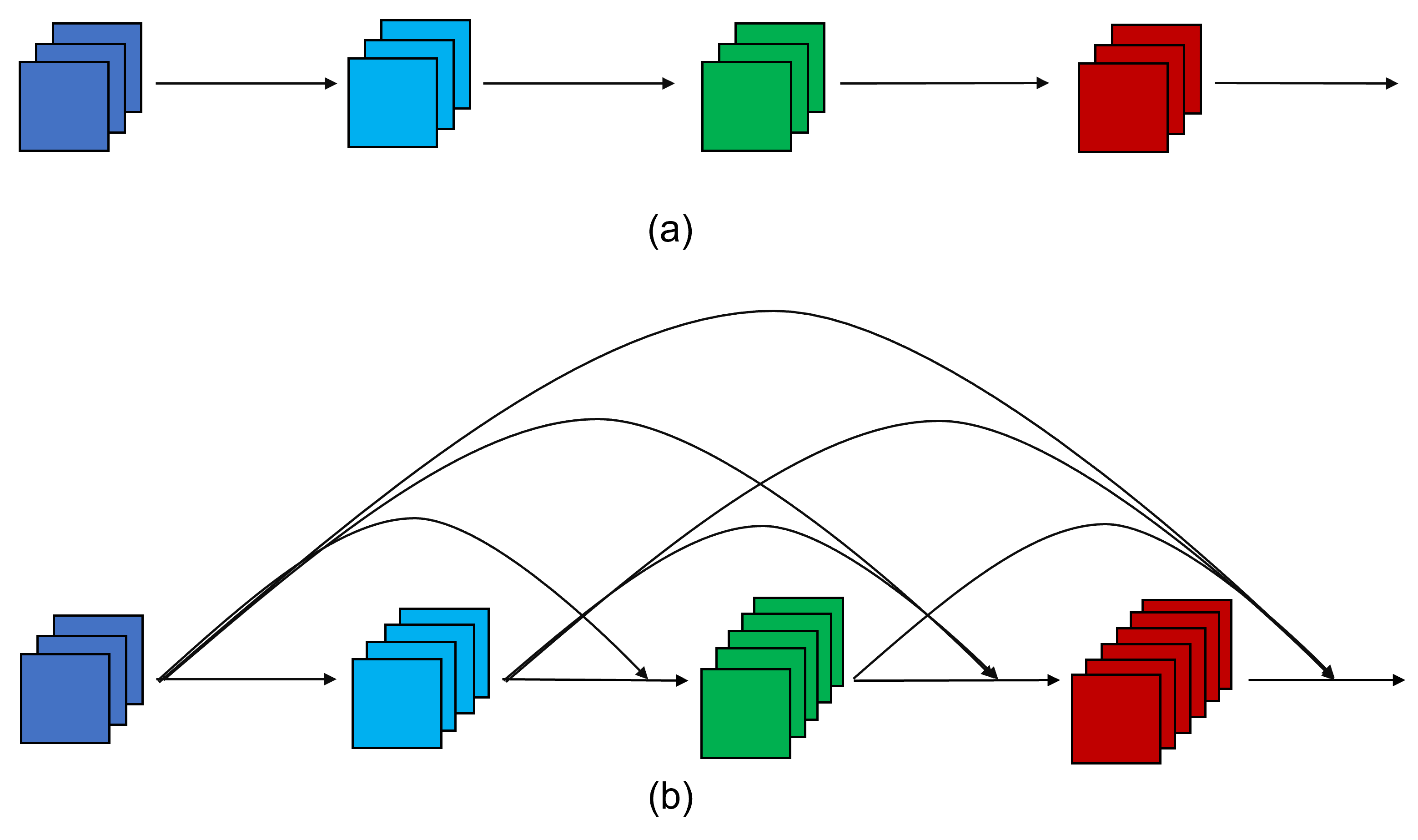}
   	\centering\caption{Difference between (a) convolutional layers and (b) densely connected convolutional layers, using a 2-D feature map as an example.}
   	\label{fig:1}
   \end{figure}

   In this work, we adopt the densely connected convolutional neural network (DCNN) \cite{huang2017densely}, an advanced CNN architecture, to learn the spatial information from the 3-D CFD data. Compared to traditional CNN, DCNN consists of multiple dense blocks of convolutional layers; within each block, there are direct connections from any layer to all subsequent layers. Such a configuration allows better information and gradient flow between the layers of the network, so that training is easier compared to a traditional CNN. The difference between CNN and DCNN is depicted in Figure \ref{fig:1}.
   
\begin{figure}[h]
   	\centering\includegraphics[width=1.0\linewidth]{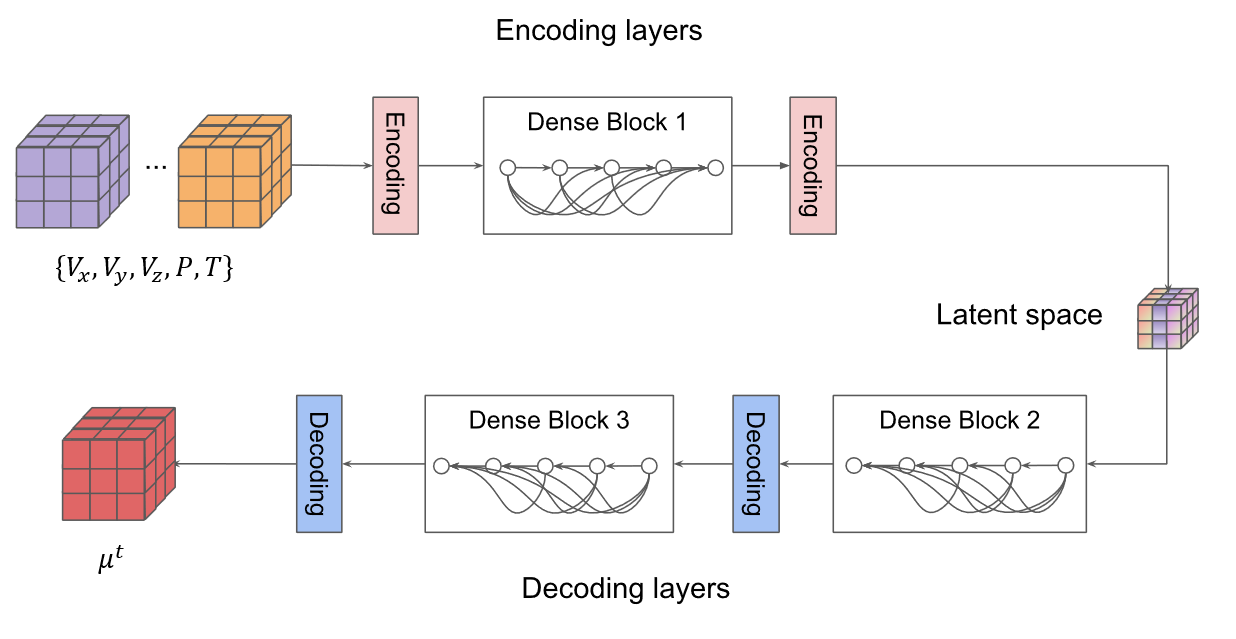}
   	\centering\caption{DCNN structure for coarse-mesh turbulent viscosity prediction.}
   	\label{fig:2}
   \end{figure}

   The designed DCNN network for spatial-information learning takes five flow-feature fields as inputs, i.e., three velocity components $V_x,\ V_y,\ V_z$, pressure $p$, and temperature $T$. The output of the network is the turbulent viscosity field $\bm{\mu}^t$. Inspired by a successful application in porous media flow \cite{zhu2018bayesian}, we adopted a similar DCNN structure that consists of an odd number of dense blocks along with an encoder-decoder layer setup, as depicted in Figure \ref{fig:2}.
   
   The encoding/decoding layer consists of two convolutional layers connected by a rectified linear unit (ReLU) layer. In the encoding layer, the first convolutional layer has $(k, s, p)$ = (1,1,0), whose purpose is to limit the number of feature maps for a more concise network. The second convolutional layer has $(k, s, p)$ = (3,2,1), which will reduce the dimensionality of the feature map by a factor of 2. The decoding layer has a similar setup, with the difference that the second convolutional layer is a transpose-convolution operation that will increase the dimensionality of the feature map by a factor of 2, sot that the ultimate output has the same dimensionality as the input flow fields. We also applied a dropout layer after each convolutional layer to serve as a regularization term to mitigate the overfitting issue \cite{gal2016dropout}. The dropout layer will randomly zero some of the elements of its input tensor with probability p, using samples from a Bernoulli distribution during the training process. Meanwhile, the outputs are also scaled by a factor of $1/(1-p)$.
   
\begin{figure}[h]
   	\centering\includegraphics[width=0.5\linewidth]{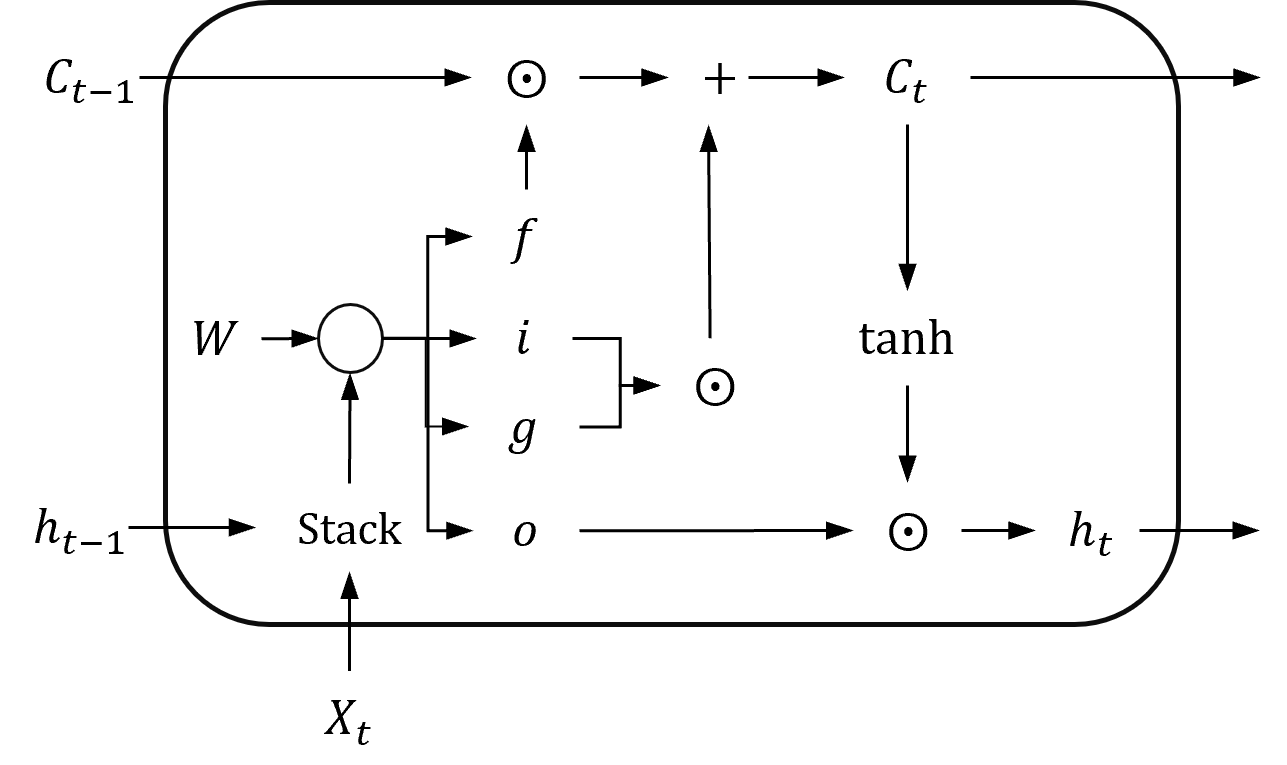}
   	\centering\caption{Information flow in a LSTM unit.}
   	\label{fig:3}
   \end{figure}

   Besides the spatial information, the 3-D CFD data for loss-of-flow transients consist of temporal information. The widely used neural-network architecture for temporal data learning is the recurrent neural network (RNN). In RNN, the result of time step $t$ depends not only on input  $x_t$ at time step $t$, but also on the hidden state $h_{t-1}$ at the previous time step. The long short-term-memory (LSTM) network is a modern RNN architecture \cite{LSTM} that has been widely used for temporal data learning. The LSTM network contains a memory cell and multiple gates to control the information flow of the cell, as depicted in Figure \ref{fig:3}. 

   In Figure 3, $f$ stands for the forget gate, which determines whether to erase the memory cell; $i$ stands for the input gate, which determines whether to write the results into the memory cell; $o$ stands for the output gate, which determines whether to export the current results stored in the memory cell; and $g$ stands for the gate that couples with the input gate to determine the specific value stored in the memory gate. The information flow within the LSTM unit is computed together with an extended trainable weight matrix $W$, and through a different non-linear activation function:
   
   \begin{equation}
   \label{eq:3}
   \begin{matrix}f_t=\sigma\left(W_{xf}\ast X_t+W_{hf}\ast\mathcal{L}_{t-1}+W_{cf}\odot C_{t-1}\right)\\\begin{matrix}i_t=\sigma\left(W_{xi}{\ast X}_t+W_{hi}\ast\mathcal{L}_{t-1}+W_{ci}\odot C_{t-1}\right)\\g_t=tanh(W_{xg}{\ast X}_t+W_{hg}\ast\mathcal{L}_{t-1})\\\end{matrix}\\\begin{matrix}o_t=\sigma\left(W_{xo}\ast X_t+W_{ho}\ast\mathcal{L}_{t-1}+W_{co}\odot C_t\right)\\C_t=f_t\odot C_{t-1}+i_t\odot g_t\\\mathcal{L}_t=o_t\odot t a n h(C_t),\\\end{matrix}\\\end{matrix},
   \end{equation} 

   \noindent where $\sigma$ stands for sigmoid activation function and $\odot$ stands for matrix elementwise multiplication.
   
   In this work, we designed a novel neural network architecture DCNN-LSTM based on the concept of convolutional recurrent neural networks \cite{xingjian2015convolutional}. Our design aims to efficiently learn the spatial-temporal CFD data with a combination of DCNN and LSTM. Based on the DCNN structure, the LSTM units take the latent space $\mathcal{L}$ obtained from DCNN as inputs; the information flow follows the one depicted in Figure \ref{fig:3}, but with a series of convolution operations instead of matrix multiplication:
   
   \begin{equation}
   \label{eq:4}
   \begin{matrix}f_t=\sigma\left(W_{xf}\ast X_t+W_{hf}\ast\mathcal{L}_{t-1}+W_{cf}\odot C_{t-1}\right)\\\begin{matrix}i_t=\sigma\left(W_{xi}{\ast X}_t+W_{hi}\ast\mathcal{L}_{t-1}+W_{ci}\odot C_{t-1}\right)\\g_t=tanh(W_{xg}{\ast X}_t+W_{hg}\ast\mathcal{L}_{t-1})\\\end{matrix}\\\begin{matrix}o_t=\sigma\left(W_{xo}\ast X_t+W_{ho}\ast\mathcal{L}_{t-1}+W_{co}\odot C_t\right)\\C_t=f_t\odot C_{t-1}+i_t\odot g_t\\\mathcal{L}_t=o_t\odot t a n h(C_t),\\\end{matrix}\\\end{matrix},
   \end{equation} 
   \noindent where $\ast$ stands for convolution operations. 
   
\begin{figure}[h]
   	\centering\includegraphics[width=1.0\linewidth]{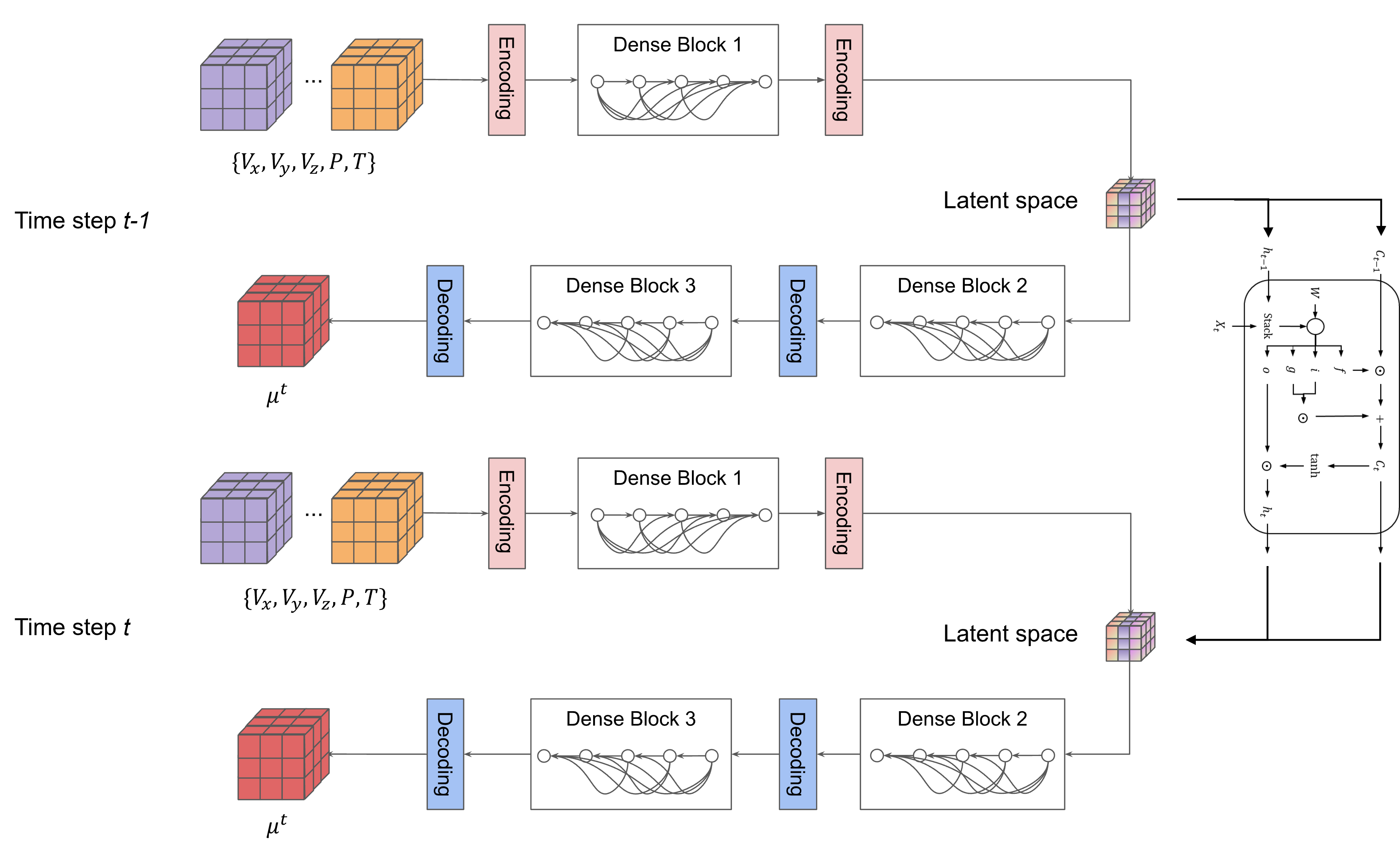}
   	\centering\caption{DCNN-LSTM architecture for coarse-mesh turbulent viscosity prediction during reactor transients.}
   	\label{fig:4}
   \end{figure}

   By taking $\mathcal{L}$ as the temporal input, DCNN-LSTM is able to learn the temporal information of the CFD transient data with a concise LSTM module, as the $\mathcal{L}$ has a compressed size of the original 3-D CFD data. The full DCNN-LSTM architecture is depicted in Figure \ref{fig:4}. Such a design ensures efficient learning for both spatial and temporal information from the 3-D transient CFD data. 
   
   \subsection{Adversarial training}
   In the normal training process for a neural network, we define a loss function $\ell(f_\theta(x),y)$ to calculate the error between neural-network predictions $f_\theta(x)$ and output data $y$. The loss function is then back-propagated through the network following the chain rule and automatic differentiation so that we can obtain the first-order derivatives of the loss function with regard to each of the neural network’s parameters. We rely on the stochastic gradient descent (SGD) algorithm to update these trainable parameters $\theta$ based on its gradient $\nabla_\theta\ell(f_\theta(x),y)$; the updating process is controlled by learning rate $\epsilon$ and batch size $B$.
   
   In the present work, the neural network model is trained with CFD data and will be used as a constitutive relation in a system code such as SAM. Numerous factors such as numerical algorithms and discretization can cause discrepancies between CFD and system code that will impact the performance of the neural network model. For this reason, the robustness of the neural network model needs to be considered to ensure that a small perturbation in the inputs will not significantly change the output of the neural network model. Therefore, we implemented the adversarial training technique \cite{goodfellow2014explaining} in addition to the SGD algorithm when training the neural network model. The fundamental idea of adversarial training is to generate “adversarial input”-output pairs and incorporate them into the training process. The adversarial input is obtained by adding a perturbation to the original input along a direction in which the network is likely to increase the loss the most. The implementation of the adversarial training is detailed in Table \ref{tab:1}.
   
   \begin{table}[h]
   	\centering
   	\caption{Adversarial training algorithm}\vspace{-3mm}
   	\label{tab:1}
   		\begin{tabularx}{\textwidth}{X}
   		    \hline
   			For each minibatch $B$:\\
   			\hspace{7mm}1. Initialize gradient vector $g$ = 0 \\
   			\hspace{7mm}2. For each input-output pair $(x,y)$ in $B$:\\
   			\hspace{14mm}a.	Calculate gradient based on original data pair\\
   			\hspace{19mm}$g=g+\mathrm{\nabla}_\theta\ell(f_\theta(x),y)$\\
   			\hspace{14mm}b. Generate adversarial input by approximately maximizing \\
   			\hspace{14mm}purturbation error\\
   			\hspace{19mm}$\delta^\star=\underset{\norm{\delta} \le \epsilon}{\argmax}(f_\theta(x+\delta^\star),y)$\\
   			\hspace{14mm}c. Add gradient at adversarial input-output pair $(x+\delta^\star,y)$\\
   			\hspace{19mm}$g=g+\mathrm{\nabla}_\theta\ell(f_\theta(x+\delta^\star),y)$\\
   			\hspace{7mm}3. Update parameters $\theta$ based on learning rate $\varepsilon$\\
   			\hspace{14mm}$\theta=\theta\ -\ \varepsilon g/\abs{B}$\\
   		    \hline
   	    \end{tabularx}
   \end{table}
   
   In the present work, we assumewe assume the results between system code and CFD can have a 5\% discrepancy, so the value of $\varepsilon$ is chosen to be 0.05. We tested the above-mentioned adversarial training process in a simplified 2-D case and confirmed that it can generate smoother prediction compared with the classical SGD process \cite{liu2021dnn_uq}. The DCNN-LSTM model is developed and trained using PyTorch \cite{pytorch}, an open-source deep learning library.
   
   \section{Problem description}
   \label{S:3}
   The performance of the proposed DCNN-LSTM model needs to be evaluated with transient simulations. In this work, we train the model with high-resolution CFD simulation results to ensure accuracy.
   
   \subsection{Loss-of-flow transients}
   The case study investigated in this work is based on the protected loss-of-flow (PLOF) transients, a key scenario in safety evaluation of a SFR design. The PLOF transient assumes the complete loss of forced flow due to multiple component failures. SFRs are typically designed with high inherent safety, and can maintain sufficient core cooling even in this extreme scenario, where the reactor-protection system is assumed to have functioned and shut down the core immediately, with only the decay heat remaining to be removed. All pumps and all emergency-powered safety systems are assumed to fail. This leaves, as the only heat removal path, the emergency heat-removal system, for which flow is maintained by natural convection alone. Any other intermediate loops, balance of plant, etc., are not assumed to provide any heat-removal capability.
   
\begin{figure}[h]
   	\centering\includegraphics[width=0.7\linewidth]{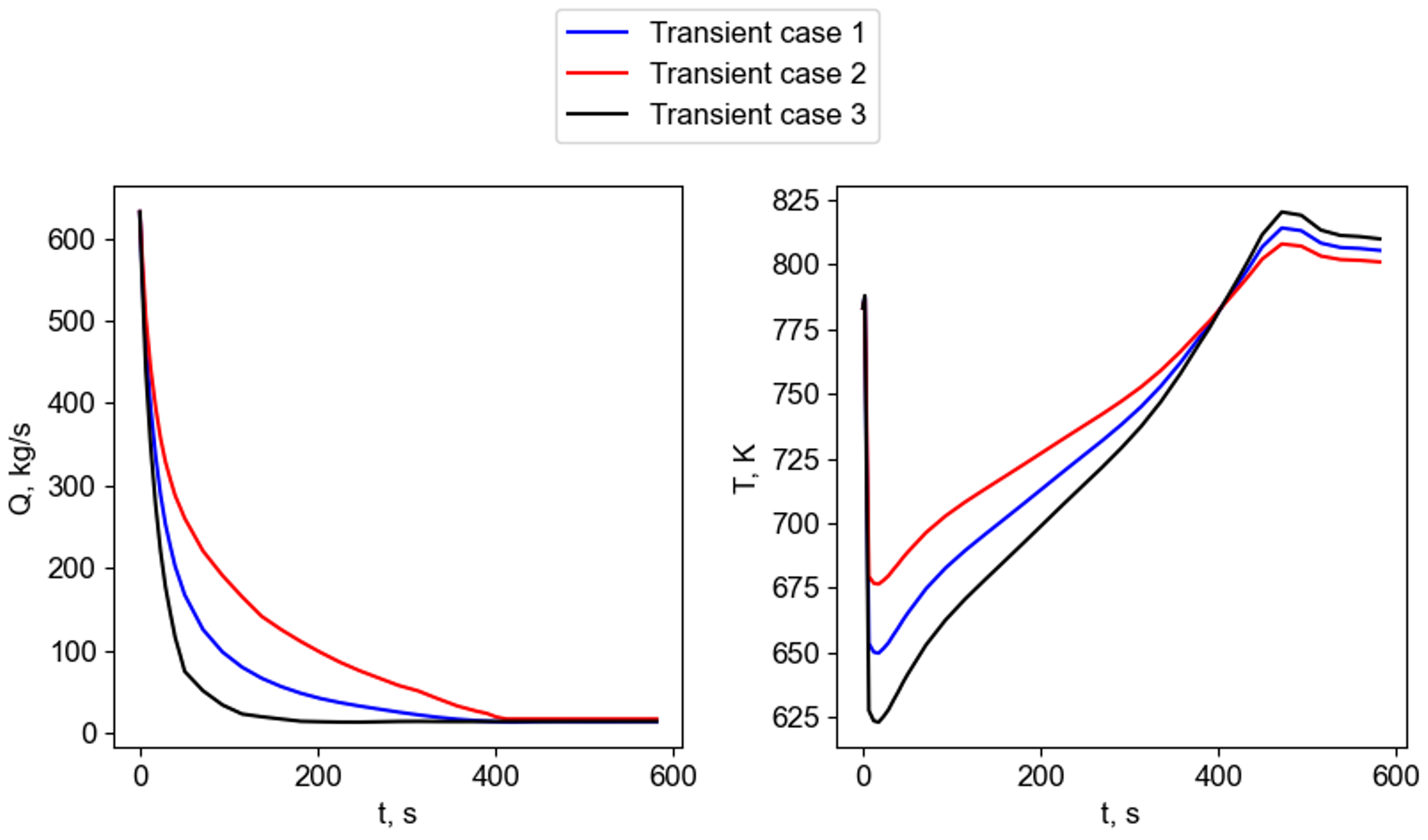}
   	\centering\caption{Three loss-of-flow transients for DCNN-LSTM training and performance evaluation.}
   	\label{fig:5}
   \end{figure}

   To make the scenario investigated here as relevant to real-world applications as possible, the transient was planned using the preconceptual design report of the Advanced Burner Test Reactor (ABTR) \cite{chang2008advanced}, a pool-type SFR design. A baseline PLOF scenario was investigated numerically for the ABTR using SAM's 1-D module. The 1-D simulation solution was used to provide inlet conditions for the CFD simulation as the baseline training transient, which is denoted as Transient 1 in the remainder of this paper. By perturbing the inlet conditions of Transient 1, two other transients, denoted as Transients 2 and 3, were simulated by CFD to test the generalization capability of the trained DCNN-LSTM model. The inlet conditions of mass flow rate $Q$ and temperature $T$ of the three transients are depicted in Figure \ref{fig:5}. Compared to Transient 1, Transient 2 has a slower mass flow rate change and a smaller temperature change, while Transient 3 has a faster mass flow rate change and a larger temperature change.
   
   \subsection{CFD simulation}
   The computation domain of the CFD simulation is a simplified cylindrical tank that represents the ABTR hot pool, as the flow mixing and thermal stratification in this region is of high importance to the safety of the SFR. The tank’s diameter is 1.95 m and its height is 7.74 m. The tank inlet is centered at the bottom origin, with a diameter of 0.43 m; its outlet is located at a height of 4.1 m, and is a narrow ellipse representing the inlet to a heat exchanger. The outlet window is 20 cm high and 2.175 m wide. It was assumed that there were two heat exchanger loops on opposite sides of the tank, resulting in a symmetrical problem. This characteristic was exploited in the simulation by using a symmetry plane along the center of the domain, opposite the outlet. The outlet was also extruded to help prevent backflow. The inlet is flush with the bottom of the tank. At steady state, the inlet mass flow rate is 632 kg/s and the inlet temperature is taken to be 783.15 K, representing half of the primary system mass flow rate and the core outlet temperature. 
   
\begin{figure}[h]
   	\centering\includegraphics[width=0.7\linewidth]{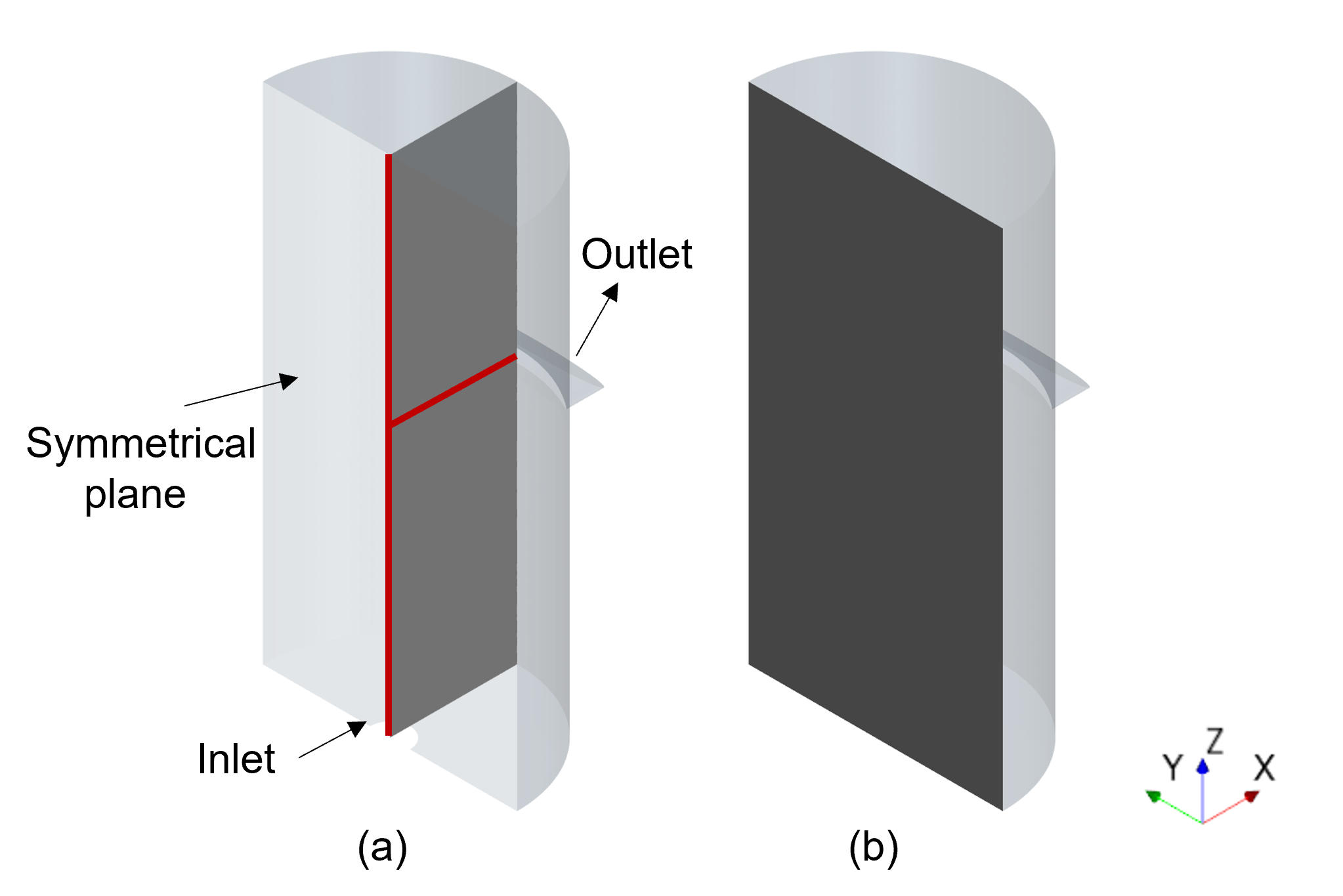}
   	\centering\caption{Computation domain of half symmetrical cylinder with data extracted for DCNN-LSTM performance evaluation at (a) outlet plane and two lines; (b) symmetrical plane.}
   	\label{fig:6}
   \end{figure}

   The computation domain is depicted in Figure \ref{fig:6}. When evaluating the performance of the DCNN-LSTM model, we extract CFD results to make qualitative and quantitative comparisons with the model’s prediction. In this work, the results at two planes were extracted for qualitative comparison, while the results at two lines (one vertical line along the vertical center of the domain, one horizontal line pointing to the outlet of the domain) were extracted for quantitative comparison. 
   
   The commercial finite-volume CFD code STAR-CCM+ \cite{star-ccm+} was used to solve the transient problem. Sodium was the working fluid, with temperature-dependent properties. Gravity acts in the –Z direction, i.e., in the opposite direction from the inlet flow. The inlet conditions of mass flow rate and temperature during the transient as described in Figure \ref{fig:5} were implemented as tables with linear interpolation between points 
   
   The top of the tank was treated as a slip wall. In an actual reactor, there is a cover gas, typically argon, to allow for space to account for changes in the sodium volume and pressure. This gas was not included, in part to allow for simpler physics for the ML model. However, it is noted that multiphase simulations of the tank were also performed using a Volume-of-Fluid model. No substantial changes in the overall flow behavior were encountered, primarily just a small raising of the interface in the center of the domain in the vicinity of the core outlet jet, and impacts on the outlet temperature were small. Thus the slip-wall simplification was deemed acceptable for the current work. 
   
   A mesh of roughly 1.05 million trimmed hexahedral cells was employed, with refinement in roughly the inner half of the tank to better resolve the jet emerging from the core. Two prismatic layers were used at the wall. A 2nd-order segregated solver was used for the flow and energy equations. The RANS-based Realizable k-$\varepsilon$ Two-Layer model was used to model turbulence in the system. This model was used along with the thermal-stratification model for the buoyancy production of dissipation. The adopted k-$\varepsilon$ turbulence model represents a compromise between accuracy and robustness, and has been shown to be reasonably accurate in prior simulations that featured buoyant jets and erosion of stratified layers \cite{kraus2015erosion}. A Reynolds-Stress model was also tested, but was shown to be similar enough to the base turbulence model that it did not warrant the additional cost. A turbulent Prandtl number of 2 was employed, which is common in simulations of liquid-metal flows. The turbulent viscosity $\mu^t$ that was computed from the obtained turbulent kinetic energy $k$ and turbulent dissipation rate $\varepsilon$ was then used as the training target of the DCNN-LSTM model:
   
   \begin{equation}
   \label{eq:5}
   \mu^t\ =\ \rho C_\mu\frac{k^2}{\varepsilon},
   \end{equation} 
   
   \noindent where $\rho$ is the density of the liquid sodium, which is a function of temperature; and $C_\mu$ is a coefficient with the fixed value of 0.09.
   
   Since a PLOF transient is assumed to start from nominal conditions, a steady-state run was performed to obtain the initial flow and temperature fields for all three transients. After that period, a variable time stepper was used to keep the mean Courant–Friedrichs–Lewy condition below 0.5, with a maximum time step of 0.1 s and a maximum change factor of 2 per time step. The transients were run for 600 s. To ensure that the time-sequence data have equal weight in training the model, we extracted the simulation results with a uniform time step of 0.2 s, so for a 600 seconds transient, there are 3000 results. As the DCNN-LSTM model considers both temporal and spatial information from the data, we further combine the 3000 snapshots into 300 samples, with each sample consisting of a consecutive sequence of 10 time steps. The DCNN-LSTM is then trained with these samples.
   
   \section{DCNN-LSTM model trainining and optimization}
   \label{S:4}
   \subsection{Data preprocessing}
   The original CFD case used more than 1.05 million nonuniform meshes for simulation. The physical quantity fields defined on such a high-resolution nonuniform mesh cannot be directly used for training the proposed DCNN-LSTM model, for two reasons. First, nonuniform mesh means the data from each mesh cell do not have equal weight, making it difficult to train a neural network model that treats each element with equal weight. Second, the high-resolution setup is not compatible with the desire for a constitutive relation for a coarse-mesh 3-D module in system code. To overcome these two issues, the k-Nearest-Neighbor (kNN) algorithm provided by the open-source ML library scikit-learn \cite{scikit} was utilized to convert the original CFD results to uniform coarse-mesh data. In the conversion, we considered only the cylindrical tank, and the outlet extrusion was not included in the converted data.
   
\begin{figure}[h!]
   	\centering\includegraphics[width=0.7\linewidth]{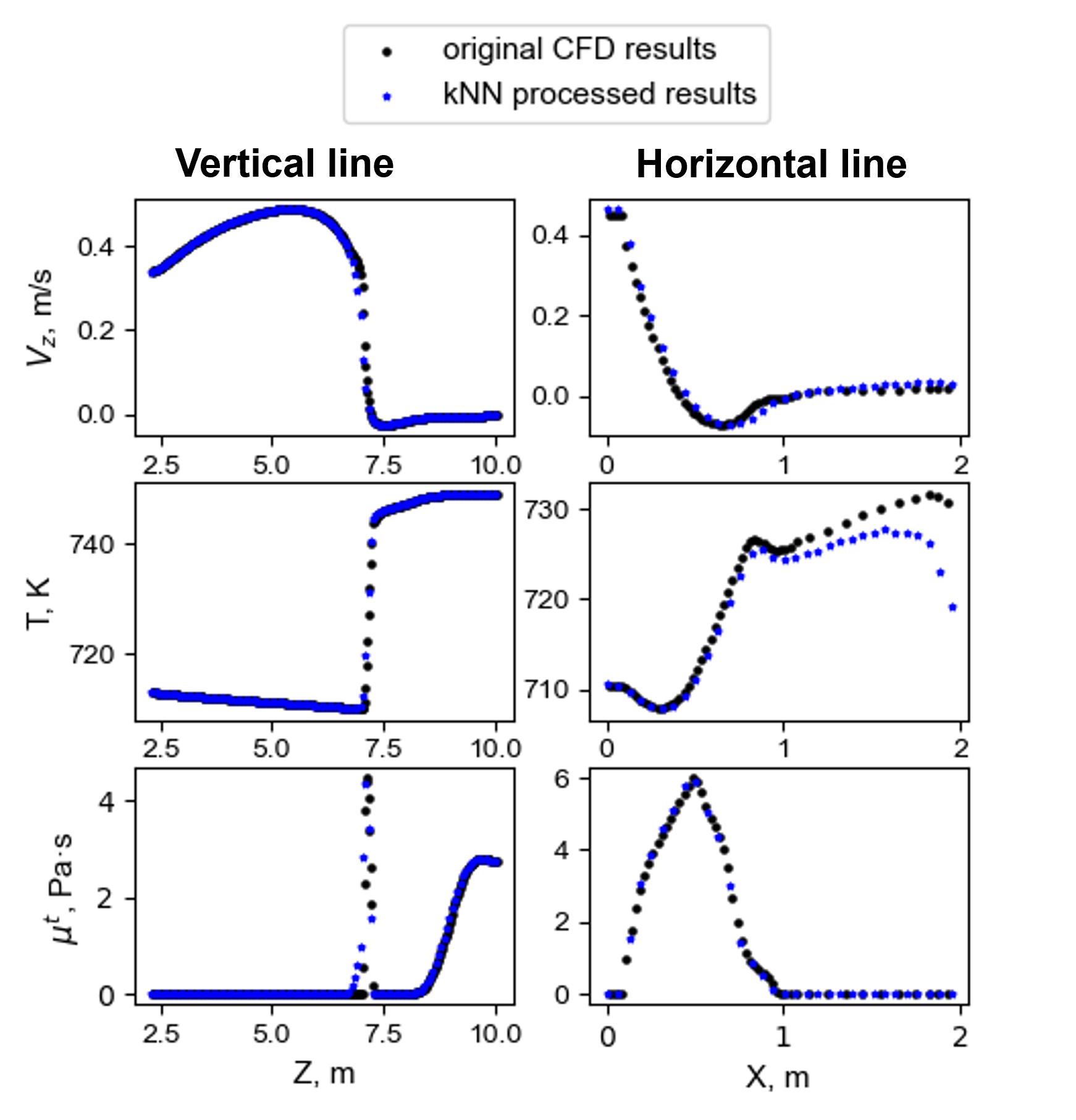}
   	\centering\caption{Quantitative comparison between original CFD results and processed kNN results, $t$ = 200 s.}
   	\label{fig:7}
   \end{figure}

\begin{figure}[h!tbp]
	\centering\includegraphics[width=0.6\linewidth]{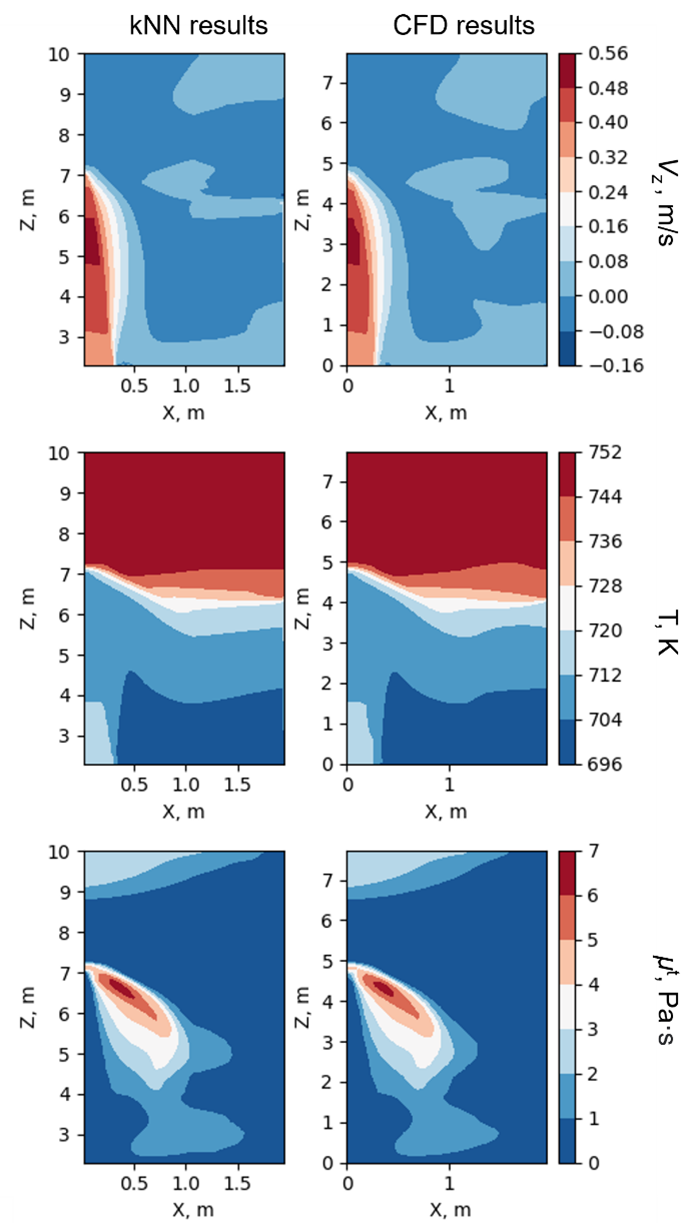}
	\centering\caption{Qualitative comparison between original CFD results and processed kNN results at outlet plane, $t$ = 200 s.}
	\label{fig:8}
   \end{figure}

   Given a point $A$ in the new uniform coarse-mesh setup, the kNN algorithm is able to find a predefined number of points in the nonuniform fine mesh closest in distance to $A$, and then predict the value of $A$ through a distance-weighted average with the values of these neighboring points. We confirmed the validity of this data preprocessing procedure by qualitatively and quantitatively comparing the original CFD results and the processed kNN results. An example of three key physical quantities, i.e., the vertical velocity component $V_z$, temperature $T$, and turbulent viscosity $\mu^t$ at $t$ = 200.0 s, is depicted in Figures \ref{fig:7} and \ref{fig:8}. The kNN results are found to be in very good agreement with the original CFD results, with the largest discrepancy ($<$ 10 K) observed in the temperature in the horizontal line pointing to the outlet center. Such an observation confirms that the data accuracy can be preserved when converting the fine-mesh nonuniform setup to a coarse-mesh uniform one.
   
   The DCNN-LSTM model takes 3-D PyTorch tensors as input and generates a 3-D tensor as output with the same dimension. In this work, the processed kNN results were padded with zero outside the semi-cylinder computation domain so that each physical scalar field can be converted to a regular 3-D tensor. Combining the kNN and padding, we were able to convert the fine-mesh nonuniform CFD results to coarse-mesh uniform tensors with dimensions of 32×64×128 for further model training.
   
   \subsection{Baseline model training}
   
   With the coarse-mesh training data ready, the next step is to obtain a baseline DCNN-LSTM model. In this work, we chose to train the model with data from Transient 1 only, while leaving the results of the two other transients as test cases to evaluate the model’s generalization capability for “extrapolating” inlet conditions. For the results of Transient 1, we further decomposed the full dataset into two parts: a training dataset with 70\% of the full dataset, and a testing dataset with the remaining 30\%. The DCNN-LSTM model was trained with the training dataset, using an error function $\ell(f_\theta(x),y)$ based on the root mean square error (RMSE) along with an additional regularization term to mitigate the overfitting of the model:
   
   \begin{equation}
   \label{eq:6}
   \ell(f_\theta(x),y)=\sqrt{\frac{1}{m}\sum_{i=1}^{m}{({\hat{y}}_i-y_i)}^2}+\ \lambda\bm{w}^T\bm{w},
   \end{equation} 
   
   \noindent where $\bm{w}$ is the learnable weights of the model, and $\lambda$ is the weight decay factor.
   
\begin{figure}[h!tbp]
   	\centering\includegraphics[width=0.7\linewidth]{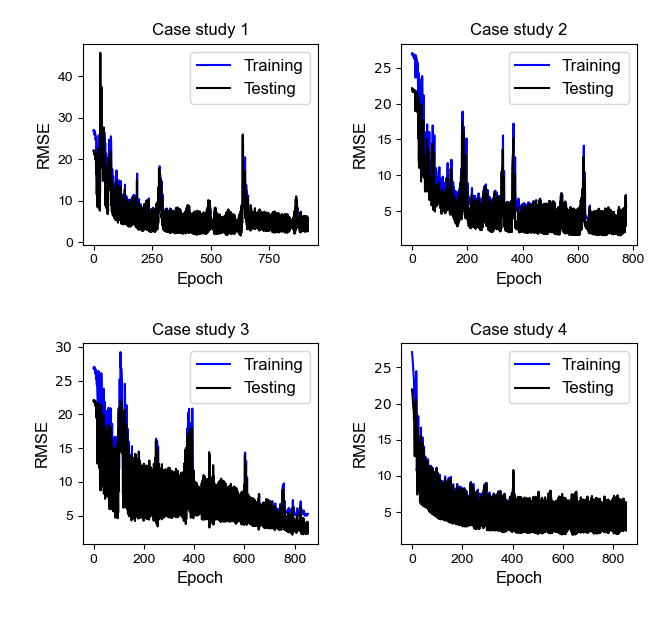}
   	\centering\caption{Training DCNN-LSTM model with different hyperparameter setups.}
   	\label{fig:9}
   \end{figure}

   Training a ML model involving LSTM module is usually more challenging than the normal feedforward network. Furthermore, it is well-known that the hyperparameters of a neural network can have a significant influence on its performance. In this work, we performed a comprehensive hyperparameter tuning process for a DCNN-LSTM model with optimal performance. The first step is to select a baseline model with reasonably low RMSE on the testing dataset by manually perturbing a few key hyperparameters, including learning rate, regularization parameter, and batch size. The RSMEs of a few case studies are depicted in Figure \ref{fig:9}.
   
   It can be seen in Figure \ref{fig:9} that the RMSEs in all cases show a fluctuating pattern when decreasing with the progress of training. Such a fluctuating pattern is typical in training the LSTM model. We also observe (especially in case study 3) that the RMSE of the training dataset is slightly larger than that of the testing dataset, as the adversarial training procedure generated “adversarial inputs” that can have larger RMSEs. We selected case study 3 as the baseline model, since it demonstrates a less fluctuating pattern with reasonably low RMSE.
   
   \subsection{Hyperparameter optimization}
   With the selected baseline model, a more comprehensive hyperparameter optimization was performed on a larger number of hyperparameters using DeepHyper \cite{balaprakash2018deephyper}, an open-source scalable automated ML package.
   
   Following the Bayesian optimization framework, DeepHyper uses an asynchronous model-based search to obtain an optimized set of hyperparameters with minimal RMSE on the testing dataset. In DeepHyper, a surrogate model $S$ that takes the hyperparameter set as input and RMSE as output is constructed. This surrogate model is an approximation of the true RMSE on the whole hyperparameter space. For a hyperparameter set $s$, $S$ predicts the mean $\mu(s)$ and standard deviation $\sigma(s)$ of the RMSE. In this work, we chose DeepHyper’s built-in random forest algorithm to construct the surrogate model, as it demonstrated superior performance compared to random search and genetic algorithm-based search \cite{maulik2020time}. Furthermore, the random-forest algorithm can handle discrete parameters (such as batch size) conveniently.
  
\begin{figure}[h!tbp]
   	\centering\includegraphics[width=1.0\linewidth]{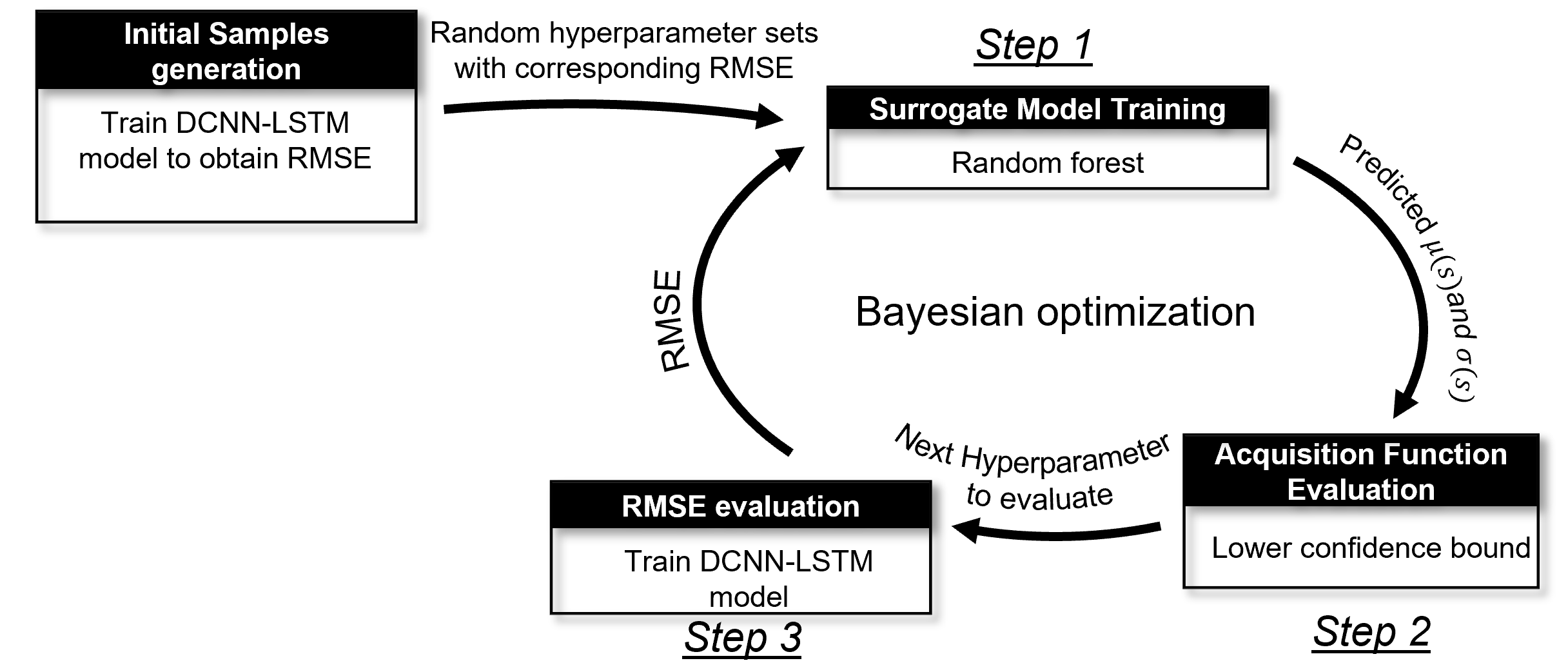}
   	\centering\caption{Bayesian optimization using DeepHyper.}
   	\label{fig:10}
   \end{figure}

   An acquisition function can be defined on the basis of $\mu(s)$ and $\sigma(s)$ for the Bayesian optimization. In this work, we chose the lower-confidence-bound acquisition function: 
   
   \begin{equation}
   \label{eq:7}
   A(s)\ =\ \mu(s)\ -\ \beta\sigma(s),
   \end{equation} 
   
   \noindent where $\beta>0$ is a tradeoff parameter. Starting from a prior surrogate model $S_{prior}$ based on a number of initial hyperparameter samples, DeepHyper iteratively identifies new samples to update $S$, so the approximation to the true RMSE on the hyperparameter space can be improved. The intuition behind the adopted acquisition function involves weighing the relative importance of the surrogate’s mean and uncertainty to achieve a balance between exploitation and exploration of the hyperparameter space that is controlled by $\beta$. In this work, we chose $\beta=1.96$ to encourage exploration. The whole Bayesian optimization process is illustrated in Figure \ref{fig:10}.
   
\begin{figure}[h!tbp]
   	\centering\includegraphics[width=0.8\linewidth]{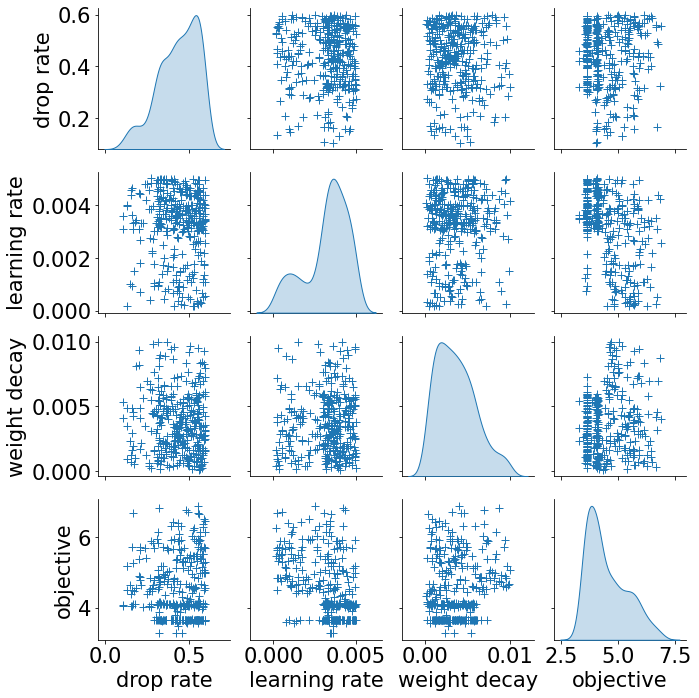}
   	\centering\caption{Pairwise plots and marginal distributions of DCNN-LSTM hyperparameter search using DeepHyper.}
   	\label{fig:11}
   \end{figure}

   Considering the relatively large computational cost for training hundreds of neural networks, we performed the Bayesian optimization with 20\% of the training data and ran for only 100 epochs. The optimization was performed with nine hyperparameters, i.e., batch size, dropout rate, growth rate of dense block, initial features of dense block, learning rate, weight-decay factor, and the number of dense layers in each of the three dense blocks. A total of 246 evaluations were performed on the Argonne Leadership Computing Facility's Theta cluster. Through the Bayesian optimization, we found that three hyperparameters have the most significant impact on the RMSE of the testing dataset, i.e., the dropout rate, learning rate, and weight decay. The pairwise plots and the marginal distributions of these three hyperparameters are depicted in Figure \ref{fig:11}.
   
\begin{figure}[h!tbp]
   	\centering\includegraphics[width=0.8\linewidth]{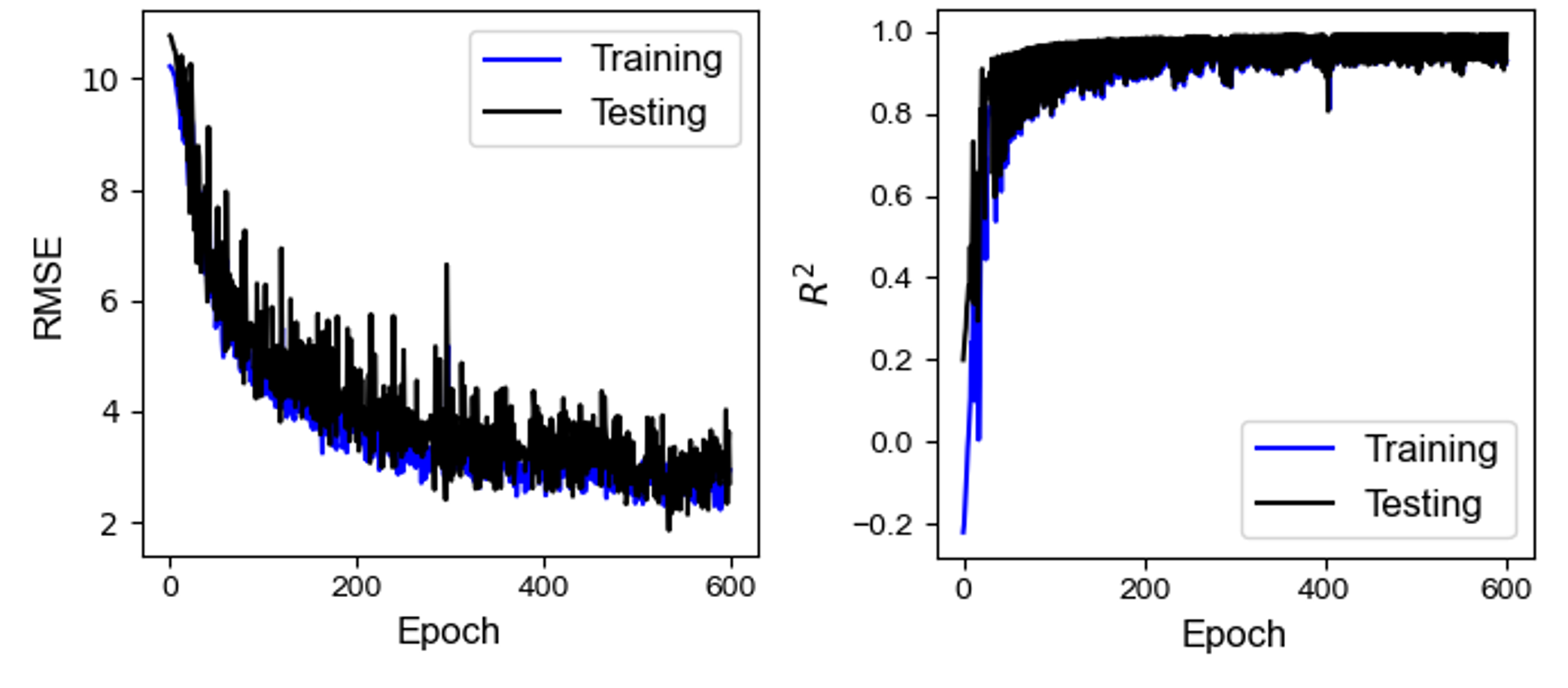}
   	\centering\caption{RMSE and R-square of optimized DCNN-LSTM training.}
   	\label{fig:12}
   \end{figure}

   With the optimized hyperparameters, we then trained the DCNN-LSTM model on the full training dataset with 600 epochs. Besides RMSE, the R-square value is also calculated to evaluate the performance of the optimized model; this value describes the ratio between the variance explained by the DCNN-LSTM model and the total variance of the dataset, and can be computed as follows:  
   
   \begin{equation}
   \label{eq:8}
   R^2=\ 1-\frac{\sum_{i=1}^{m}{({\hat{y}}_i-y_i)}^2}{\sum_{i=1}^{m}{({\bar{y}}_i-y_i)}^2},
   \end{equation} 
   
   \noindent where ${\bar{y}}_i$ is the mean value of the output. An $R^2$ value close to 1 indicates very good regression of the model. As demonstrated in Figure \ref{fig:12}, the optimized DCNN-LSTM model has better performance compared to the baseline model. 
   
   \section{Performance evaluation}
   \label{S:5}
   The performance evaluation of the optimized DCNN-LSTM model comprised two separate parts. In the first part, the DCNN-LSTM model was evaluated on the testing dataset of Transient 1, which consists of 30\% of the full dataset. In the second part, the DCNN-LSTM model was evaluated on Transients 2 and 3, whose results were not used in training the model. The purpose of the first part is to evaluate the effectiveness of the training and optimization process, while the purpose of the second part is to evaluate the generalization capability of the model on transients with “extrapolating” inlet conditions. 
   
   \subsection{Evaluation of the training effectiveness of DCNN-LSTM}
\begin{figure}[p]
   	\centering\includegraphics[width=0.65\linewidth]{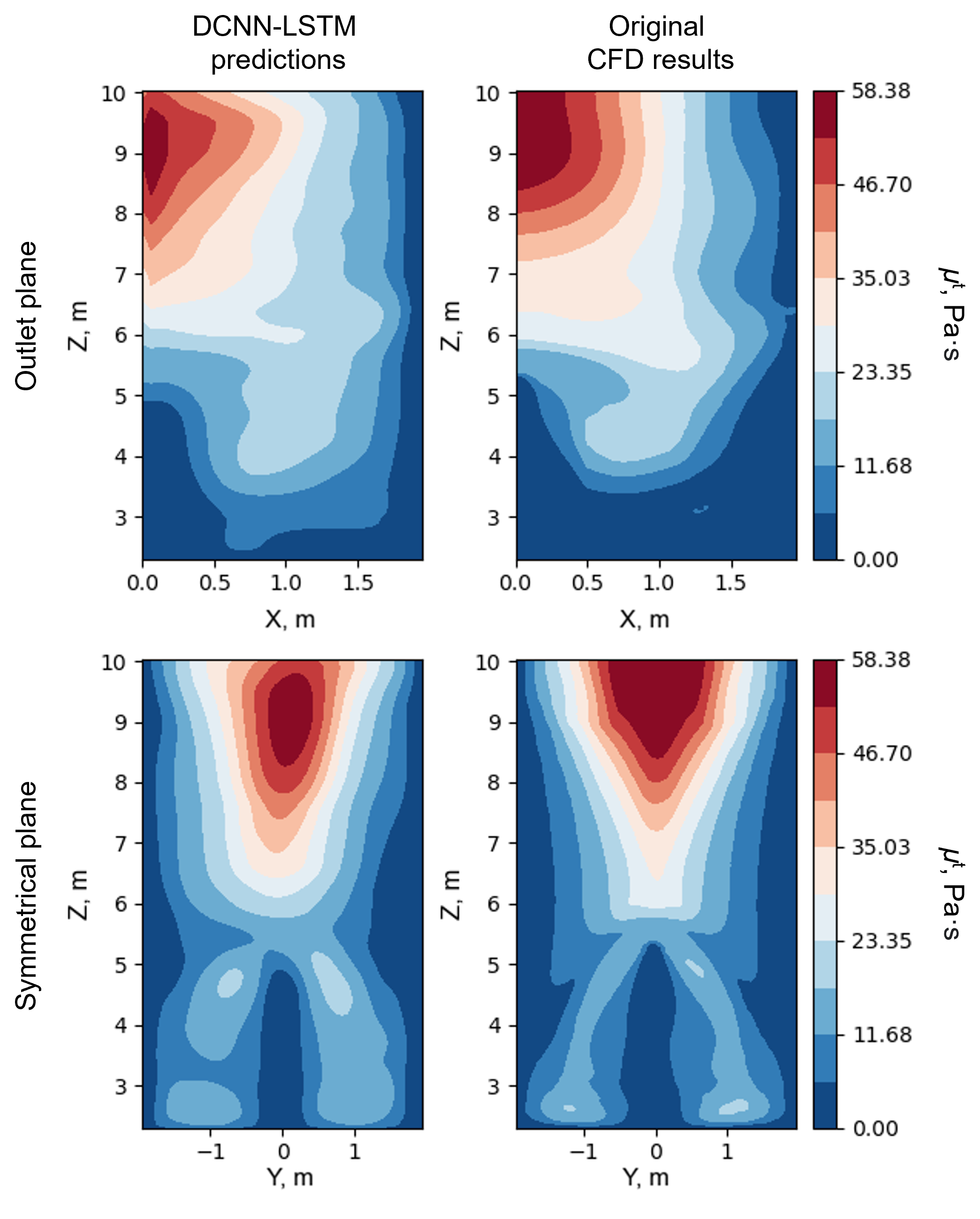}
   	\centering\caption{Qualitative comparison between DCNN-LSTM prediction and original CFD results at $t$ = 70.2 s, Transient 1.}
   	\label{fig:13}
   \end{figure}

\begin{figure}[p]
	\centering\includegraphics[width=0.7\linewidth]{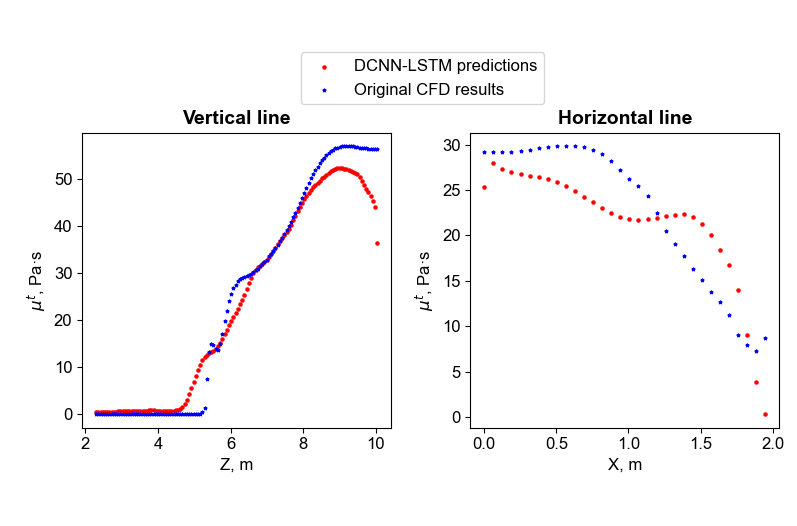}
	\centering\caption{Quantitative comparison between DCNN-LSTM prediction and original CFD results at $t$ = 70.2 s, Transient 1.}
	\label{fig:14}
   \end{figure}

\begin{figure}[p]
	\centering\includegraphics[width=0.65\linewidth]{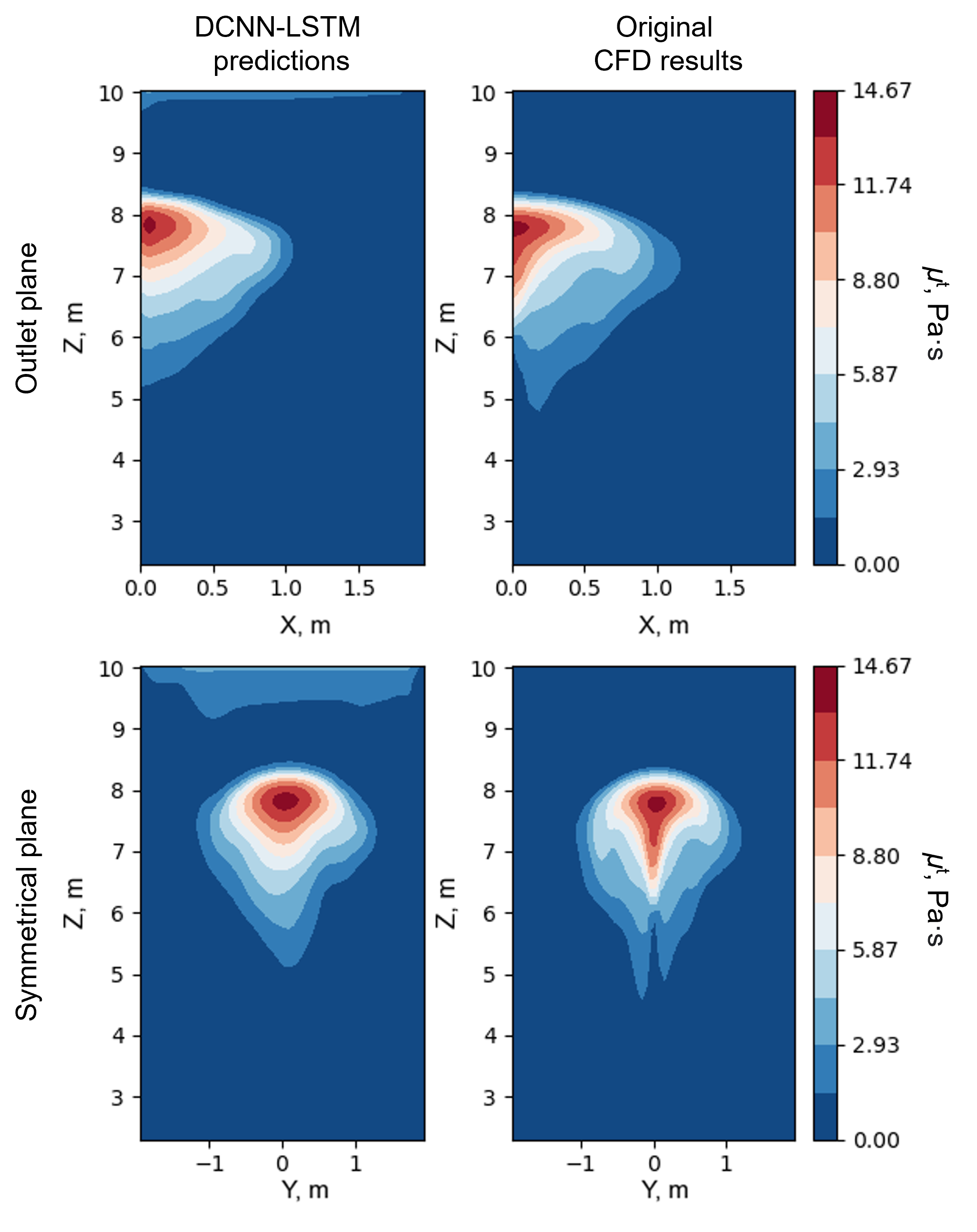}
	\centering\caption{Qualitative comparison between DCNN-LSTM prediction and original CFD results at $t$ = 400.2 s, Transient 1.}
	\label{fig:15}
   \end{figure}

\begin{figure}[p]
   	\centering\includegraphics[width=0.7\linewidth]{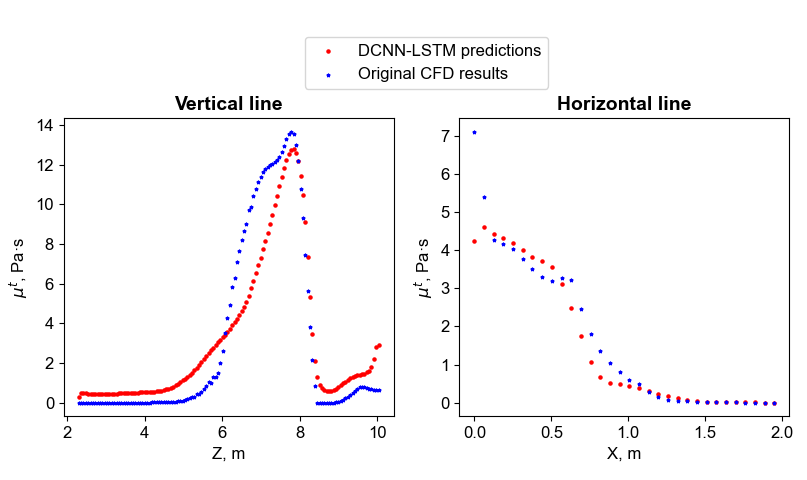}
   	\centering\caption{Quantitative comparison between DCNN-LSTM prediction and original CFD results at $t$ = 400.2 s, Transient 1.}
   	\label{fig:16}
   \end{figure}

   The DCNN-LSTM model takes sequences of the five physical fields, i.e. $V_x,\ V_y,\ V_z,\ p,$ and $T$, to predict the sequence of turbulent viscosity field $\mu^t$. The results of two snapshots at $t$ = 70.2 s and 400.2 s are extracted and compared with the original CFD results. Both qualitative and quantitative results are depicted in Figures \ref{fig:13}-\ref{fig:16}. 
   
\begin{figure}[p]
   	\centering\includegraphics[width=0.8\linewidth]{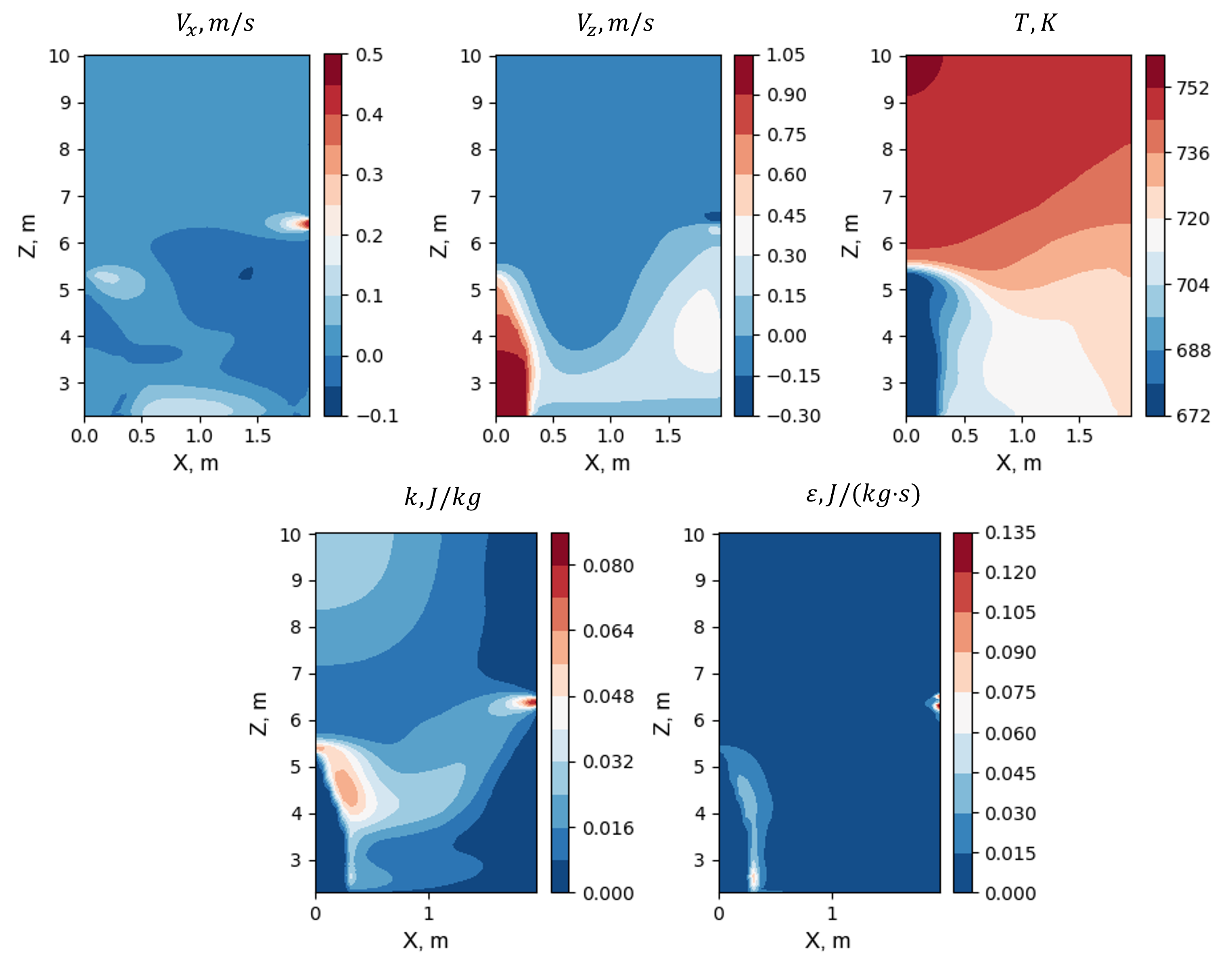}
   	\centering\caption{CFD results for major physical quantities on outlet plane at $t$ = 70.2 s, Transient 1.}
   	\label{fig:17}
   \end{figure}

\begin{figure}[p]
	\centering\includegraphics[width=0.8\linewidth]{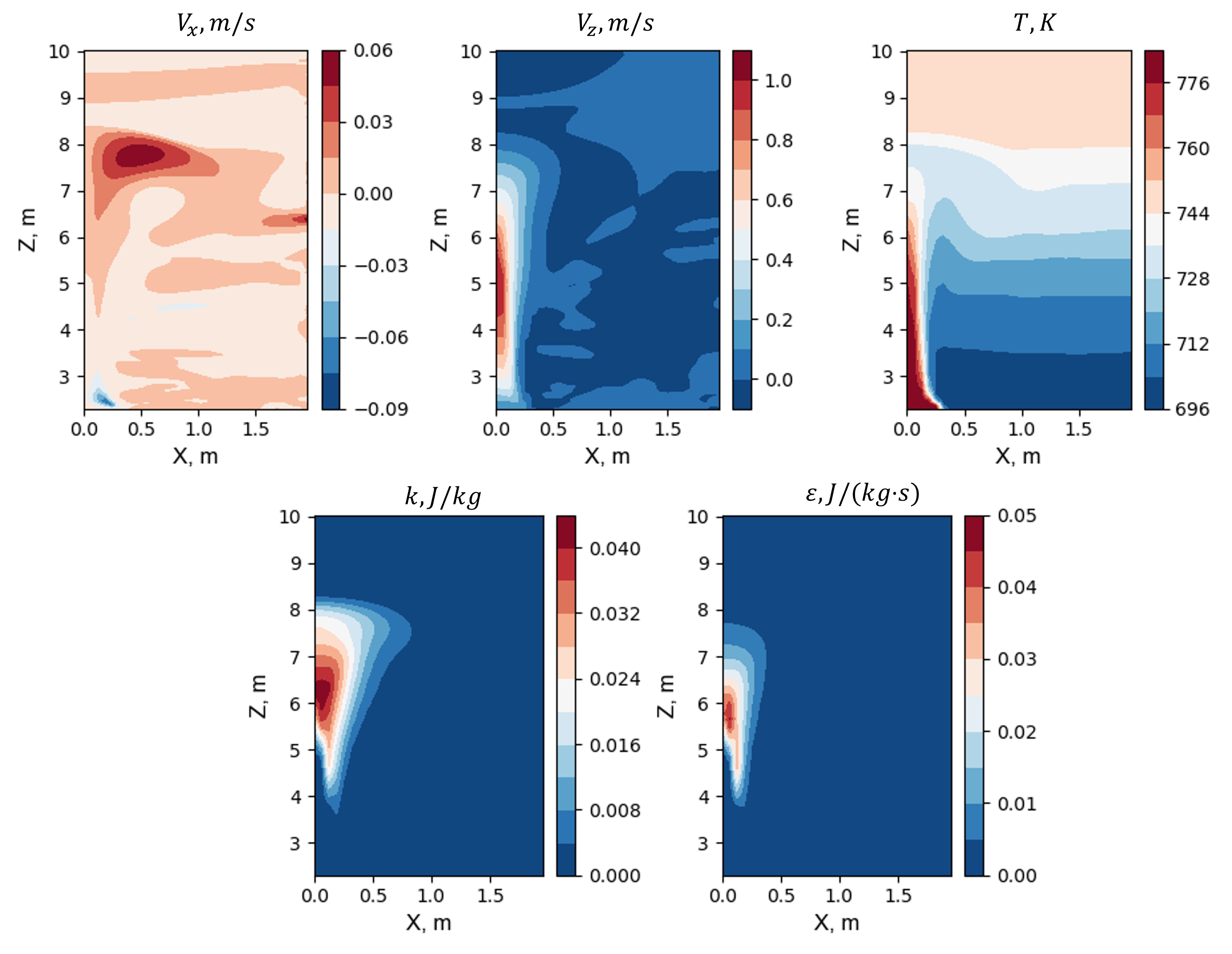}
	\centering\caption{CFD results for major physical quantities on outlet plane at $t$ = 400.2 s, Transient 1.}
	\label{fig:18}
   \end{figure}

   It can be found from the four figures that the turbulent viscosity showed a very dynamic pattern change during the progress of the transients. Such a dynamic change is anticipated, as we observe dynamic changes in the inlet mass flow rate and temperature during the transient. The observed turbulent viscosity fields are also consistent with the major physical-quantity fields obtained from the CFD simulation at these two time snapshots, as depicted in Figures \ref{fig:17} and \ref{fig:18}. As can be seen from these two figures, at $t$ = 70.2 s, the thermal stratification is still forming, the inlet mass flow rate is at 20\% of its initial value, and the inlet sodium is colder than its initial temperature. At this time step, the turbulent kinetic energy $k$ is high at the edge of the marching cold liquid sodium, the outlet region, and the center top region, while the turbulent dissipation rate $\varepsilon$ is high at the edge of the marching liquid sodium and the outlet region. As the product of $k$ and $\varepsilon$, the turbulent viscosity is high in the top center region of the pool. At $t$ = 400.2 s, the thermal stratification is stabilized and the inlet mass flow rate is only 2\% of the initial value. The inlet sodium is much hotter than the liquid sodium stabilized in the bottom region of the pool, thus creating a buoyancy force to drive the hot sodium upward. Both $k$ and $\varepsilon$ are high at the edge of the marching hot sodium, thus creating a turbulent viscosity field that is high in the region of the rising hot sodium. 
   
   It can also be found that the optimized DCNN-LSTM model can capture the dynamic pattern change. Overall, we can conclude that the DCNN-LSTM model demonstrated high accuracy in predicting the testing dataset of Transient 1. The observation confirms the effectiveness of the training and optimization process. On the other hand, we can observe that the DCNN-LSTM model does not resolve the sharp turbulent viscosity change well, and tends to predict a smoother turbulent viscosity field. This behavior can be caused by the kernel size we apply and the information loss in the decoding/encoding process. Furthermore, we can also observe in the 400.2 s snapshot that the DCNN-LSTM model overestimated the turbulent viscosity in the top region of the pool. This can be caused by the residual memory stored in the LSTM module, as at the early stage of the transient the turbulent viscosity is high in this upper region. We also observe a slightly more asymmetric pattern in the DCNN-LSTM prediction compared to the original CFD data. This difference can be caused by the additional error introduced in the kNN conversion process.
      
   \subsection{Evaluation of the generalization capability of DCNN-LSTM}
   
   Compared to the effectiveness of training the DCNN-LSTM model, it is more important to evaluate the generalization capability of the trained network with cases of “extrapolating” inlet conditions, as the ultimate purpose of training the neural network model is to apply it to a variety of conditions that do not have available CFD simulations. In this work, the generalization capability of the DCNN-LSTM model is tested with two other loss-of-flow transients with the inlet conditions perturbed from Transient 1, as shown in Figure \ref{fig:5}. 
   
\begin{figure}[p]
   	\centering\includegraphics[width=0.65\linewidth]{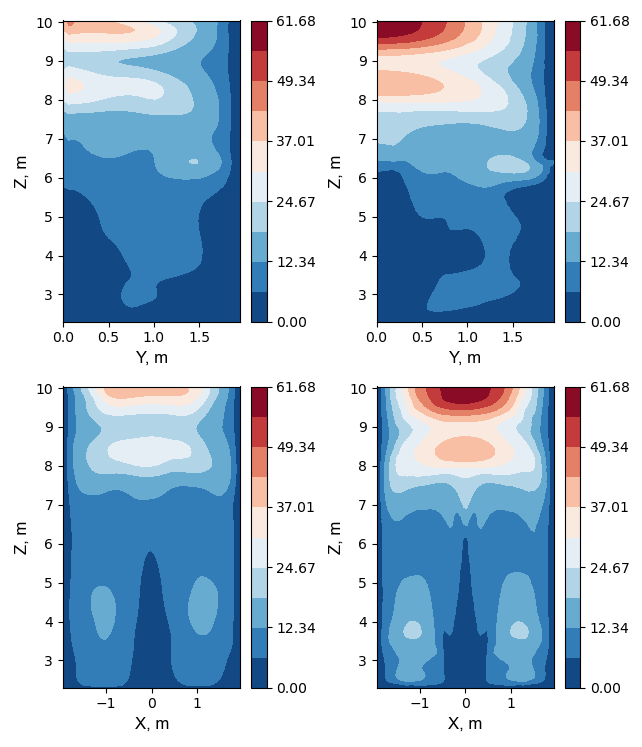}
   	\centering\caption{Qualitative comparison between DCNN-LSTM prediction and original CFD results at $t$ = 70.2 s, Transient 2.}
   	\label{fig:19}
   \end{figure}

\begin{figure}[p]
   	\centering\includegraphics[width=0.7\linewidth]{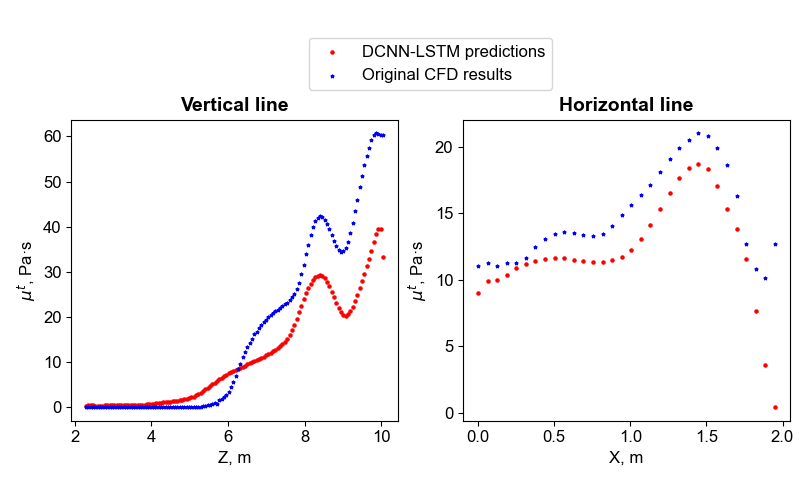}
   	\centering\caption{Quantitative comparison between DCNN-LSTM prediction and original CFD results at $t$ = 70.2 s, Transient 2.}
   	\label{fig:20}
   \end{figure}

\begin{figure}[p]
   	\centering\includegraphics[width=0.65\linewidth]{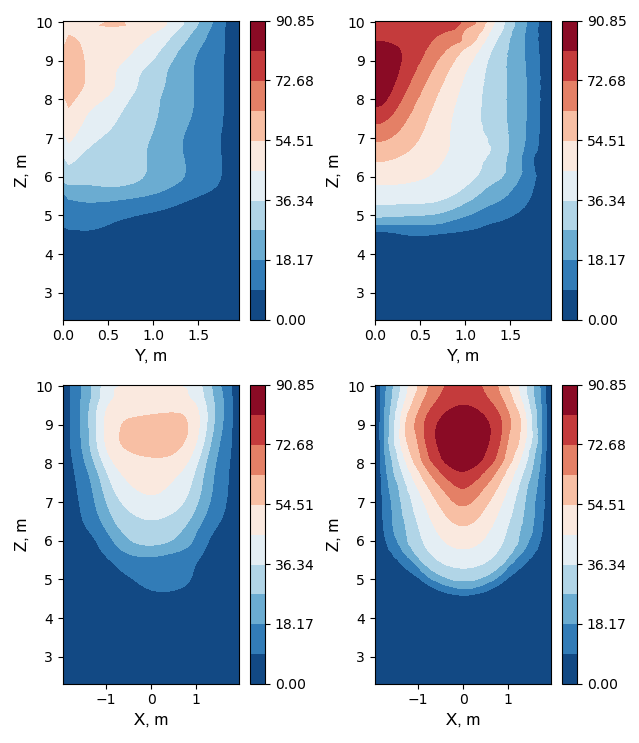}
   	\centering\caption{Qualitative comparison between DCNN-LSTM prediction and original CFD results at $t$ = 70.2 s, Transient 3.}
   	\label{fig:21}
   \end{figure}

\begin{figure}[p]
   	\centering\includegraphics[width=0.7\linewidth]{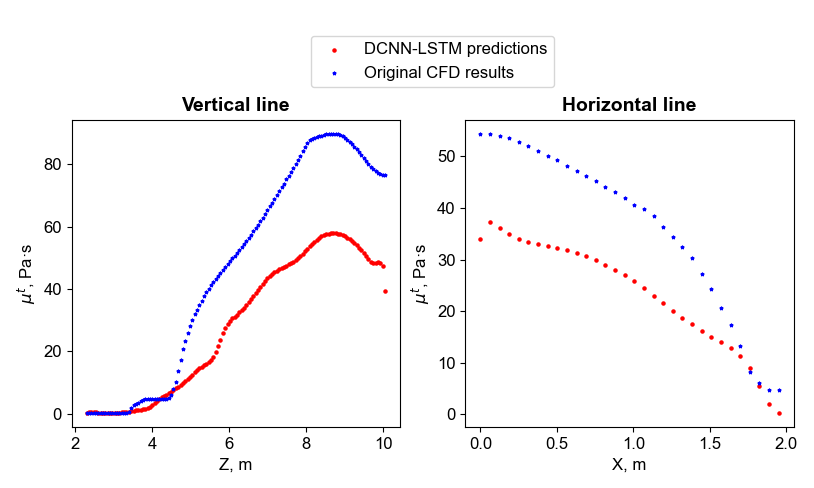}
   	\centering\caption{Quantitative comparison between DCNN-LSTM prediction and original CFD results at $t$ = 70.2 s, Transient 3.}
   	\label{fig:22}
   \end{figure}

   In a similar manner to the training effectiveness evaluation, we obtain the sequence of the turbulent viscosity field for Transients 2 and 3. The snapshots at $t$ = 70.2 s of both cases are used for comparison. The qualitative and quantitative comparisons are depicted in Figures \ref{fig:19}-\ref{fig:22}.
   
   We can observe from the figures that compared to the results of Transient 1 at the same time step, Transients 2 and 3 showed distinctive turbulent viscosity fields, suggesting that a moderately perturbed inlet condition can cause significant differences in local flow fields. It can also be found that the turbulent viscosity field predicted by the DCNN-LSTM model is generally consistent with the CFD results in terms of trend and distribution. However, the prediction underestimated the turbulent viscosity for both cases. The generally consistent turbulent viscosity field indicates the model has generalization capability for “extrapolating” inlet conditions. But the underestimation suggests that the generalization capability is not good enough for the model to give predictions on “extrapolating” cases at the same accuracy as the training case.
   
   \subsection{Explanation of DCNN-LSTM’s generalization capability}
   
   We employed t-distributed stochastic neighbor embedding (t-SNE) \cite{van2008visualizing} to study the intrinsic data similarity among the CFD results of the three transients. The t-SNE method is a popular method for exploring high-dimensional data; it deals with data in high-dimensional space and find a faithful representation of those data in a lower-dimensional space, typically 2-D or 3-D for visualization purposes. Through t-SNE, the data with high similarity are mapped into nearby points and data with high dissimilarity are mapped into distant points with high possibility. In this work, we use the scikit-learn library to perform the t-SNE.
   
\begin{figure}[h]
   	\centering\includegraphics[width=1.0\linewidth]{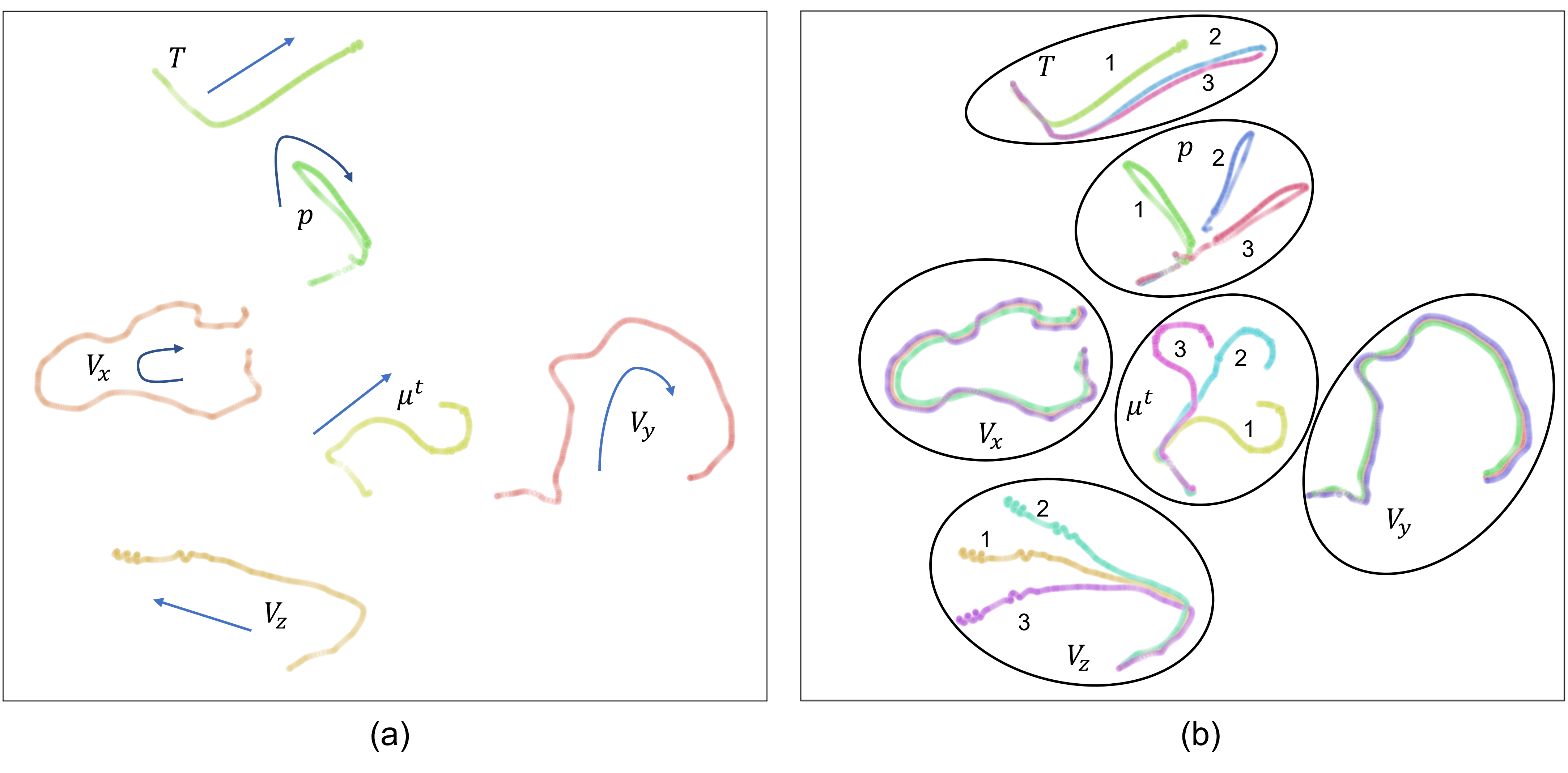}
   	\centering\caption{t-SNE mapping of physical quantity fields at each time step: a) Transient 1 only (arrow direction indicates the progress of transient); b) all three transients.}
   	\label{fig:23}
   \end{figure}

   We used t-SNE to map the scalar fields of six physical quantities (five inputs and one output) into a 2-D map. To eliminate the scaling influence on the t-SNE map, all the physical quantities are normalized by the mean and standard deviation of the transient they belong to. Each physical field at a time snapshot is a 3-D tensor with dimensions $32\times64\times128$. These high-dimensional data are converted to a point in the 2-D t-SNE map for visualization purpose. The resulting maps are depicted in Figure \ref{fig:23}. 
   
   It can be found from Figure \ref{fig:23}(a) that the six physical quantities of Transient 1 can be clearly distinguished in the 2-D t-SNE map, suggesting fundamental intrinsic differences among them. Furthermore, the six quantities at each time step form six line structures that have clear directions as the transient progresses. Such an observation indicates the temporal continuity of these physical quantities during the transient and confirms the necessity to use the LSTM module to learn the temporal information.
   
   The physical quantities of all three transients are plotted together in Figure \ref{fig:23}(b), from which we can clearly identify six clusters, with each corresponding to a physical quantity. As observed for Transient 1, the physical quantities of Transients 2 and 3 also form lines with directions in accordance with the progress of the transient. It can be further found from Figure \ref{fig:23}(b) that every physical quantity of these three cases has a high similarity at the beginning of the transient, as they are very close to each other within each cluster. This similarity diminishes as the transient progresses, except for $V_x$ and $V_y$. As observed from the t-SNE map, most physical quantities of the three cases at the end of the transient are much less similar compared to the beginning. The possible reason for the high similarity of $V_x$ and $V_y$ through the whole transient is that both velocity components lack a specific driving source; instead, they are driven by the inlet sodium in the Z direction. As a result, the values of these two fields are relatively small over the whole domain, so the t-SNE map does not distinguish the difference among the three transients. 
   
   Such an observation provides an explanation of DCNN-LSTM’s performance on Transients 2 and 3: The model has moderate generalization capability for these two transients, as their data have some level of intrinsic similarity with the training data of the model (Transient 1); however, such a similarity is not great enough to fully cover the whole transients, so the model showed a less accurate prediction for Transients 2 and 3 compared to its prediction for Transient 1. Be that as it may, the performance of DCNN-LSTM on Transients 2 and 3 still provides promise that we can leverage a well-trained data-driven model for simulation cases with “extrapolating” inlet and boundary conditions.  
   
   There are two potential approaches to improving the performance of a DNN model such as DCNN-LSTM for cases with “extrapolating” inlet/boundary conditions. The first is to train the model with more transients that cover a variety of inlet/boundary conditions, so that the trained DCNN-LSTM model can generalize over a broad range of local flow features. In this way, for a case with “extrapolating” inlet/boundary conditions, the simulation results can find similarities with the training data more easily. Such a solution is backed by the finding that for a data-driven model trained with local flow features such as DCNN-LSTM, the extrapolation of global inlet and boundary conditions does not necessarily lead to extrapolation of the model’s input. In pioneering research on fluid flow problems, Bao et al. successfully demonstrated several cases in which global extrapolation of inlet/boundary conditions can still lead to interpolation of local flow features. The results show that a global-scale gap was bridged in a greater significance by deeply exploring the scale invariance in local data. \cite{bao2020using, bao2020computationally}. The second approach is to embed physics directly into the data-driven model so the model can generate results that are consistent with the physical system. One example is physics-informed neural networks, which have enabled significant progress on many physical problems involving the solution of PDEs \cite{karniadakis2021physics}. Particularly, one can combine both approaches to improve the DNN model. One example is the composite neural network \cite{meng2020composite}, where multiple neural networks are trained with data of different fidelity, e.g. a small set of high fidelity DNS/LES data and a larger set of moderate-fidelity RANS data, these networks can be combined with embedded physics to achieve good generality.
   
   \section{Conclusions}
   \label{S:6}
   In this paper, we proposed a data-driven modeling approach based on deep neural networks to develop constitutive relations to support coarse-mesh multi-dimensional modeling in modern nuclear-system code. We demonstrated the applicability of this approach with a data-driven turbulence model applied during loss-of-flow transients that involve thermal mixing and stratification phenomena in a pool-type sodium-cooled fast reactor. The proposed turbulence model has a coarse-mesh setup to ensure computational efficiency, while it is trained with fine-mesh CFD data to ensure accuracy. We developed a novel neural network architecture termed DCNN-LSTM that combines a densely connected convolutional network and a long-short-term-memory network. DCNN-LSTM can efficiently learn the spatial-temporal information from the CFD transient-simulation results. Utilizing the fine-mesh CFD simulation results from a loss-of-flow transient, we performed a comprehensive training and hyperparameter optimization process to obtain a DCNN-LSTM model with optimal performance. We then evaluated the generalization capability of the optimized DCNN-LSTM with two transients that have “extrapolating” inlet conditions compared to the training transient.
   
   We found that the DCNN-LSTM model can predict the turbulent viscosity with high accuracy over the whole training transient. This observation confirms the effectiveness of the training and optimization process. For the two cases with “extrapolating” inlet conditions, the DCNN-LSTM model could still capture the general distribution of the turbulent viscosity field, but showed discrepancies compared to the original CFD results. This observation indicates that the DCNN-LSTM model did have moderate generalization capability for the “extrapolating” inlet conditions. We then employed the t-distributed stochastic neighbor embedding  algorithm to study the data similarity between the training transient and the two “extrapolating” transients. We found that the physical fields of the three transients have certain similarities, but the similarity is not great enough to fully cover the whole transients, so the model is less accurate on the transients with extrapolating inlet conditions compared to the training transient. Be that as it may, the results still provide promising insights indicating that we can leverage a well-trained data-driven model for simulation cases with “extrapolating” inlet and boundary conditions.
   
   We believe there are two possible approaches to improving the generalization capability of the data-driven model that is trained with local flow features. The first is to include more transients with a variety of different inlet and boundary conditions to form a comprehensive training database, making it easier for a given “extrapolating” case to find local similarity with the training data. The second is to directly embed physics into the model, so the model can generate results that are consistent with the physical system. Future investigations on this topic will pursue these two directions.

   \section*{Acknowledgement}
   This material is based upon work supported by Laboratory Directed Research and Development (LDRD) funding from Argonne National Laboratory, provided by the Director, Office of Science, of the U.S. Department of Energy under Contract No. DEAC02-06CH11357.
   
   The authors gratefully acknowledge the computing resources provided on Blues, a high-performance computing cluster operated by the Laboratory Computing Resource Center at Argonne National Laboratory.
   
   The authors gratefully acknowledge the resources of the Argonne Leadership Computing Facility, which is a DOE Office of Science User Facility supported under Contract DE-AC02-06CH11357.






\newpage
\bibliographystyle{elsarticle-num-names}
\bibliography{DL_turbulence.bib}







\end{document}